\documentclass[a4paper, amsfonts, amssymb, amsmath, reprint, showkeys, nofootinbib, twoside, superscriptaddress]{revtex4-1}
\usepackage[english]{babel}
\usepackage[utf8]{inputenc}
\usepackage[colorinlistoftodos, color=green!40, prependcaption]{todonotes}
\usepackage[caption=false]{subfig}
\usepackage{braket}
\usepackage{comment}
\usepackage{amsthm}
\usepackage{mathtools}
\usepackage{physics}
\usepackage{xcolor}
\usepackage{graphicx}
\usepackage[left=23mm,right=13mm,top=35mm,columnsep=15pt]{geometry} 
\usepackage{adjustbox}
\usepackage{placeins}
\usepackage[T1]{fontenc}
\usepackage{lipsum}
\usepackage{csquotes}
\usepackage[pdftex, pdftitle={Article}, pdfauthor={Author}]{hyperref} 
\begin{document}
\title{Optimizing Entanglement Generation and Distribution Using Genetic Algorithms}

\author{Francisco Ferreira da Silva}
    \email{f.hortaferreiradasilva@tudelft.nl}
    \affiliation{QuTech, Delft University of Technology, Lorentzweg 1, 2628 CJ Delft, The Netherlands}
    \affiliation{Kavli Institute of Nanoscience, Delft University of Technology, Lorentzweg 1, 2628 CJ Delft, The Netherlands}
\author{Ariana Torres-Knoop}
    \affiliation{SURF Utrecht,  Postbus 19035, 3501 DA Utrecht, The Netherlands}
\author{\\Tim Coopmans}
    \affiliation{QuTech, Delft University of Technology, Lorentzweg 1, 2628 CJ Delft, The Netherlands}
    \affiliation{Kavli Institute of Nanoscience, Delft University of Technology, Lorentzweg 1, 2628 CJ Delft, The Netherlands}
\author{David Maier}
    \affiliation{QuTech, Delft University of Technology, Lorentzweg 1, 2628 CJ Delft, The Netherlands}
    \affiliation{Kavli Institute of Nanoscience, Delft University of Technology, Lorentzweg 1, 2628 CJ Delft, The Netherlands}
\author{Stephanie Wehner}
    \affiliation{QuTech, Delft University of Technology, Lorentzweg 1, 2628 CJ Delft, The Netherlands}
    \affiliation{Kavli Institute of Nanoscience, Delft University of Technology, Lorentzweg 1, 2628 CJ Delft, The Netherlands}

\date{\today} 

\begin{abstract}
 Long-distance quantum communication via entanglement distribution is of great importance for the quantum internet. However, scaling up to such long distances has proved challenging due to the loss of photons, which grows exponentially with the distance covered. Quantum repeaters could in theory be used to extend the distances over which entanglement can be distributed, but in practice hardware quality is still lacking. Furthermore, it is generally not clear how an improvement in a certain repeater parameter, such as memory quality or attempt rate, impacts the overall network performance, rendering the path towards scalable quantum repeaters unclear. In this work we propose a methodology based on genetic algorithms and simulations of quantum repeater chains for optimization of entanglement generation and distribution. By applying it to simulations of several different repeater chains, including real-world fiber topology, we demonstrate that it can be used to answer questions such as what are the minimum viable quantum repeaters satisfying given network performance benchmarks. This methodology constitutes an invaluable tool for the development of a blueprint for a pan-European quantum internet. We have made our code, in the form of NetSquid simulations and the \textit{smart-stopos} optimization tool, freely available for use either locally or on high-performance computing centers.
\end{abstract}


\maketitle

\section{Introduction} \label{sec:introduction}
A quantum internet could be used to perform tasks that are impossible with classical communications alone, the best known example being that of quantum key distribution (QKD) \cite{bennett2020quantum, ekert1991quantum}. Beyond QKD, several other applications have been identified, ranging from quantum clock synchronization \cite{komar2014quantum} to distributed quantum computing \cite{buhrman2003distributed}. The level of network development required is application-dependent, but all of them rely on entanglement generation and distribution \cite{wehner2018quantum}.

Entanglement generation has been demonstrated at short distances~\cite{hensen2015loophole}, but scaling up has proved very challenging due to the exponential growth of photon losses with the length of fiber covered. Classically, photon loss is overcome by direct amplification, but in the quantum case this is impossible for non-orthogonal states due to the no-cloning theorem. As an alternative, two distant end nodes can be connected by intermediate nodes, known as quantum repeaters \cite{briegel1998quantum}. These are devices that can, in theory, enable long-distance entanglement generation.

Despite ongoing experimental efforts, a scalable quantum repeater has yet to be demonstrated \cite{rozpkedek2019near}. Several physical systems are being explored as possible platforms for such a repeater, for example color centers in diamond (e.g. NV centers~\cite{humphreys2018deterministic}), atomic ensembles \cite{yu2020entanglement} and trapped ions \cite{zwerger2017quantum}, but it is not yet clear which are most feasible in the short-term, nor how imperfections in the physical system would generally affect relevant network performance metrics like end-to-end fidelity and entanglement generation rate. 

In the quest for a quantum internet, the question of what the minimal requirements on a quantum repeater are to achieve a certain network performance benchmark is thus fundamental. Furthermore, as we will show in Section \ref{sec:methodology}, it is a question that can be framed as an optimization problem. Broadly speaking, two different approaches are being explored in the theoretical study of quantum repeaters: analytical (see e.g.~\cite{munro2015inside, sangouard2011quantum, briegel1998quantum}) and simulation-based (see e.g.~\cite{netsquid, coopmans2020netsquid, van2008system, wu2020sequence}). In the first case, simplifying assumptions are made, such as approximating the states shared between nodes by Werner states or assuming simple topologies for the networks under study, e.g. restricting the analysis to chains of equally spaced nodes. In the second case, accurate and realistic simulations of networks of quantum repeaters are developed, at both the protocol level and the physical level. Each of these approaches offers benefits and drawbacks; opting for an analytical approach enables obtaining analytical expressions for interesting metrics such as the end-to-end fidelity. This means that traditional gradient-based optimization methods can be employed, offering a clear path to an answer. However, this may come at the cost of less detailed predictive power. On the other hand, choosing a simulation-based approach allows the modelling to be as realistic as desired. The downside here is, of course, that analytical results are no longer available, and that the simulations may become very complex, especially taking into account the exponential overhead in simulating quantum systems on classical computers. It also renders optimizing more difficult, as all that is made available to the optimization procedure are the inputs and outputs of the simulation in consideration. For example, having more information about the function landscape, such as number of minima, would render the optimization process simpler. Besides this, a realistic simulation requires a large number of parameters, thus resulting in a large search space to be explored. Nonetheless, if one wants to arrive at a realistic answer, a simulation-based approach seems inevitable.

In this work we propose a methodology based on genetic algorithms and simulations of quantum repeaters for optimization of entanglement generation and distribution in quantum networks. This allows us to answer questions such as what are the worst possible repeaters satisfying target benchmarks. Contrasting with previous work on repeater chain optimization~\cite{wallnofer2020machine, muralidharan2016optimal, jiang2007optimal, santra2019quantum, goodenough2020optimising, krastanov2019optimized}, our methodology constitutes a systematic and modular approach to this problem, successfully integrating simulation and optimization tools, as well as allowing for the use of high-performance computing clusters.  
\section{Summary of Results}
\begin{figure}[!ht]
\centering
\includegraphics[width=\columnwidth]{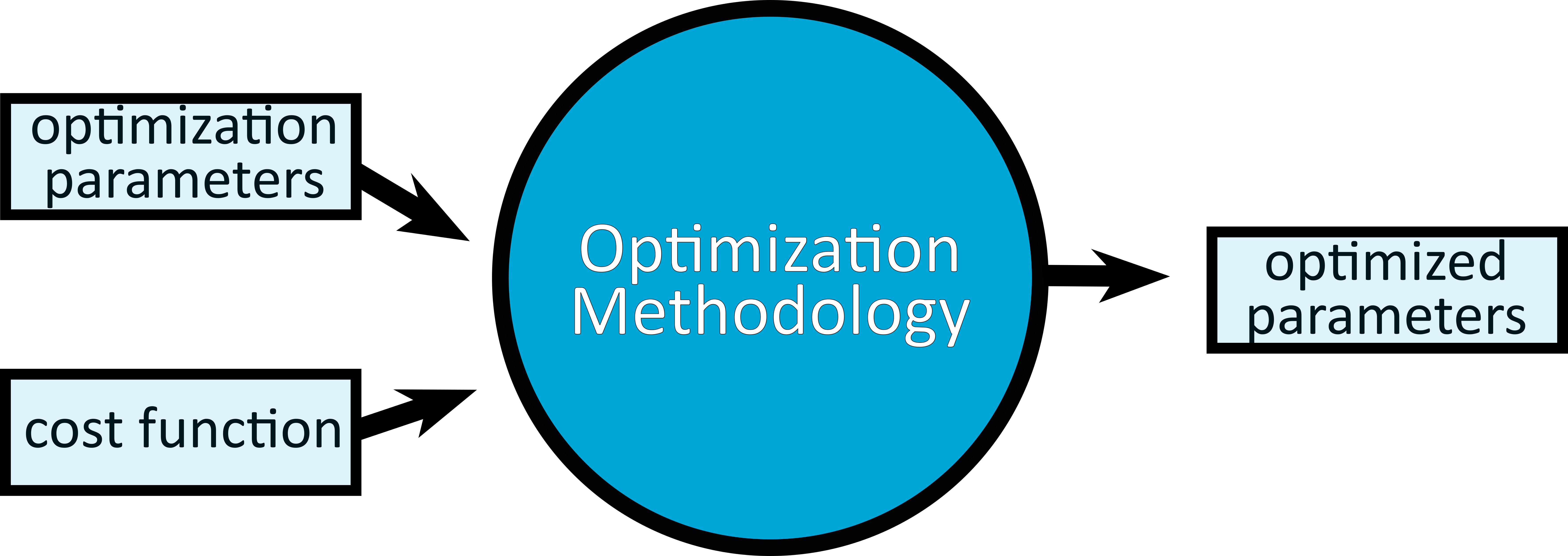}
\caption{Overview of our optimization process. The user inputs the desired optimization parameters and defines a cost function. Using simulation and optimization tools, our methodology finds a set of parameters optimizing the cost function. For example, the optimization parameters could be parameters defining a quantum repeater model and the cost function could be the inverse of the secret key rate plus a penalty term for parameter values that are much better than a given baseline. The output would then be the values of the parameters defining the quantum repeater model optimizing the cost function.}
\label{fig:results_summary}
\end{figure}

The main result of this work is the introduction of a methodology for the optimization of entanglement generation and distribution in quantum networks. Making use of genetic algorithms (GAs) and repeater chain simulations, our methodology allows for such optimizations, both locally and on high-performance computing clusters. A high-level overview of how a user interfaces with this process is shown in Figure~\ref{fig:results_summary}. 
We performed our simulations using NetSquid \cite{netsquid, coopmans2020netsquid}. NetSquid can accurately model the effects of time-dependent noise, rendering it well equipped to predict quantum network performance in a physically accurate setting. The tools used in this methodology, which allow for running NetSquid simulations together with an optimization algorithm both locally and on an HPC cluster, are made freely available (see~\cite{smartstopos}).

We structure the paper as follows. In Section~\ref{sec:methodology} we introduce our methodology, together with the required preliminaries. Section~\ref{sec:validation} concerns the validation of the methodology. This comprises two steps: (i) benchmarking our GA implementation by running it on standard optimization problems and comparing its performance to those found in the literature; and (ii), validating our approach by applying it to a repeater chain where elementary link states are in the Werner form and all noise sources are depolarizing~\cite{werner1989quantum}. In this case, analytical results can be found, so we can evaluate how well our optimization method performs. 

After validating our methodology, we apply it to some use cases in order to demonstrate its potential usefulness. We present these results in Section~\ref{sec:usecase}, where we first consider a repeater chain based on real-life fiber data, courtesy of SURF, a network provider for Dutch education and research institutions. This showcases the power of our simulation-based approach, as chains of unevenly spaced nodes are hard to study analytically. We further apply our methodology to chains of varying length, internode distance and number of repeaters and we compare the solutions found with our methodology for each of these different setups. This allows us to investigate how the impact of the parameters varies across setups, thus identifying possible bottlenecks and paths towards scalable quantum repeaters. 
\section{Methodology}
\label{sec:methodology}
In this section we introduce the main contribution of our work, a methodology for the optimization of entanglement generation and distribution. We first present each of the elements that are used in this optimization process. We finalize the section with an overview of how they are integrated to answer the question of what are the minimum requirements on quantum repeaters to achieve a given benchmark.
\subsection{Question}    
We aim to answer the question of what the minimum requirements are on the quality of quantum repeaters to achieve a given benchmark by framing it as an optimization problem. To do so, we must first clarify what we mean by requirements and by quality of a quantum repeater. Let us say that a quantum repeater is described, in a given model, by a set of $N$ parameters $\{x_i\}_{i \in \{1,...,N\}}$. The meaning of $x_j$ is model dependent. For example, if we consider a model of a  trapped ion system, $x_j$ and $x_k$ could be single-qubit and two-qubit gate error probabilities. We could also, in a more abstract model, combine these two parameters together to obtain a swap quality that quantifies the noise introduced in an entanglement swap operation, which would then be $y_j$ in this model. The quality of a quantum repeater is then a function of the set of parameters describing it. This also helps clarify what we mean by requirements. Suppose we have some fixed network topology and performance metric. To give a concrete example, the topology could be a repeater chain of $10$ equally spaced nodes and the performance metric the end-to-end secret key rate. The requirements on the repeaters are then the worst set of parameters that enable attaining some value of the end-to-end secret key rate over a chain of $10$ nodes, i.e. the lowest quality repeaters satisfying this metric.

\subsection{Cost}
\label{sec:cost}

This being said, what we mean by repeater quality is still not completely clear. To see this, let us say that we have two repeaters described by a set of parameters $\{y_i\}_{i \in \{1,...,N\}}$ and $\{z_i\}_{i \in \{1,...,N\}}$, and that the values of these parameters are the same for all but two of them, i.e. $\{y_i\} = \{z_i\}$ $\forall i \in \{1,2,\ldots,N\}\setminus \{j,k\}$. Let us further say that $y_j$ is better than $z_j$, but $z_k$ is better than $y_k$. Which of these sets of parameters is the better one? To answer this, we will now introduce the quantity to be optimized, the cost function. We emphasize that our method is completely general and could be applied to any cost function, but for concreteness we focus on a particular one from here on out.


We expect that in an experimental setting a given physical parameter becomes harder to improve the closer to its perfect value it is, so we would like our cost function to reflect this. We start by transforming our parameters so that they all live in the $[0,1]$ interval, with $1$ being the perfect value and $0$ the worst possible value. We refer to Appendix \ref{sec:appendiGA} for details. Denoting $x_b$ as the baseline value of a parameter, i.e. the value from which we are improving, $k$ as the improvement factor and $x_{\text{new}}$ as the new improved value, we claim that the following equation reflects this behaviour:

\begin{equation}
x_{\text{new}}(k) = x_b^{\frac{1}{k}}.
\label{eq:progressiveimprovement}
\end{equation}   
This can be read as: we improve $x_b$ by a factor $k$ to get $x_{\text{new}}$. To see that Equation \ref{eq:progressiveimprovement} does in fact reflect the desired behaviour, we note the that
\begin{align}
    x_{\text{new}}(k=1) &= x_b,\label{eq:factor1improv}\\
    \lim_{k \to \infty} x_{\text{new}} &= 1.\label{eq:infiniteimprovement}
\end{align}
Equation \eqref{eq:factor1improv} can be read as: improving a parameter by a factor of $1$ is equivalent to not improving it all, whereas Equation \eqref{eq:infiniteimprovement} can be taken to mean that in order to improve a parameter to its perfect value we must improve by a factor of infinity, i.e. there is no such thing as a perfect process.

We can then define the cost associated to $x_{\text{new}}$ as the factor $k$ by which we must improve the baseline value $x_b$ to obtain $x_{\text{new}}$. Therefore, solving Equation~\eqref{eq:progressiveimprovement} for $k$, we get
\begin{equation}
    k = \frac{1}{\log_{x_b}(x_{\text{new}})}.
\label{eq:improvement_factor}
\end{equation}

With this in hand, we can finally define the cost associated to a set of parameters. Let us say our model is described by a set of parameters $\{x_i\}_{i \in \{1,...,N\}}$, and that the current baseline value of each of these parameters is $\{x_{i_b}\}_{i \in \{1,...,N\}}$. A set of values $\{x_{i_c}\}_{i \in \{1,...,N\}}$ is mapped to a cost, $C$, by Equation~\eqref{eq:costfunction}. Intuitively, this can be seen as taking the average of the cost associated to each of the parameters.

\begin{equation}
    C\left(x_{1_c}, ..., x_{N_c}\right) = \sum_{i=1}^N \frac{1}{\log_{x_{i_b}}\left(x_{i_c}\right)}
    \label{eq:costfunction}
\end{equation}

There is still the matter of how the network's target performance metrics are taken into account. Throughout this work we will focus on fidelity $F$ of the end-to-end state with the ideal Bell state and entanglement generation rate $R$, but we stress that our method is not limited to optimizing for these quantities. More concretely, we will try to answer the question of what are the minimum requirements on repeaters to concurrently achieve certain values of $F$ and $R$. We are then faced with a multi-objective problem, as we want to optimize multiple quantities simultaneously, namely end-to-end fidelity, entanglement generation rate and parameter cost. Furthermore, there are trade-offs between these goals. For example, improving the memory lifetime of nodes in a chain has a positive contribution towards end-to-end entanglement fidelity, but a negative one towards parameter cost. There is a multitude of possible ways of approaching such problems \cite{gunantara2018review}. We chose to map our multi-objective optimization problem to a single-objective one by assigning weights to the different objectives and adding them, a process known as scalarization~\cite{schaffer1986some}. The total cost function $T_C$ we will minimize then becomes:

\begin{multline}
    T_C\left(p_{1_c}, ..., p_{N_c}, F_{min}, R_{min}\right) = w_1 \Theta(F_{min} - F) \\+ w_2 \Theta(R_{min} - R) +  w_3C\left(p_{1_c}, ..., p_{1_N}\right),
\label{eq:totalcost}
\end{multline}
where the $w_i$ are the weights of each objective, $\Theta$ is the Heaviside function, defined in Equation \eqref{eq:heaviside}, and $F_{min}$ and $R_{min}$ are, respectively, the minimum required end-to-end fidelity and end-to-end entanglement generation rate. Using step functions reflects the idea that we are looking for solutions that satisfy performance benchmarks, with no reward given for surpassing them. The weights in Equation \eqref{eq:totalcost} are hyperparameters of our method, meaning that they are not determined by some algorithm but must instead be chosen. This choice has an impact on which sets of parameters have the lowest costs and hence on the solutions found by the method. For example, if we assign very high values to $w_1$ and $w_2$ and a low value to $w_3$ the best sets of parameters will be those that satisfy the requirements on the end-to-end fidelity and rate without much regard for how costly it is to achieve them. The parameter cost term, defined in Equation \eqref{eq:costfunction}, depends on the baseline values of the parameters, which must also be chosen. Typically, for the use cases we consider, these will be chosen to reflect what is currently achievable experimentally.

\begin{equation}
    \Theta (x) = \left\{
\begin{array}{ll}
      0 & x < 0 \\
      1 & x \geq 0\\
\end{array} 
\right. 
\label{eq:heaviside}
\end{equation}

Optimal solutions to this single-objective optimization problem are then solutions to the multi-objective optimization problem. 

\subsection{Abstract Model}
\label{sec:abstractmodel}
In order to explore and better understand the methodology we propose, we believe it to be wise to employ a relatively simple model whose behavior we understand. We must however again emphasize that our methodology is completely general in terms of the model used for the quantum repeater hardware.

Before introducing our model, we briefly go over some basic repeater concepts. This is not meant to be an in-depth introduction to quantum repeaters; for that, we direct the reader to a review of the subject, such as~\cite{munro2015inside}. Quantum repeaters are devices that can in theory be used to connect two distant end nodes. This is done by (i) establishing elementary links between neighbouring nodes, i.e. entangled states shared by these nodes and (ii) glueing links together by means of bell state measurements, a process known as entanglement swapping. The simplest possible quantum repeater protocol consists in having nodes constantly trying to generate entanglement and swapping as soon as they hold two entangled qubits, one to each side of the chain. This is known as a SWAP-ASAP protocol. It can be enhanced by imposing a cut-off condition, such as a maximum time after which stored entanglement is discarded~\cite{collins2007multiplexed}. For simplicity, we will refrain from doing so and we will run all our simulations using a SWAP-ASAP protocol with no cut-offs.

We consider a simplified five-parameter model for a quantum repeater, the five parameters being denoted by $[F_{EL}, p_{suc}, s_q, T_1, T_2]$. We assume that elementary links states have fidelity $F_{EL}$ with the ideal Bell state and that they are generated with a success probability $p_{suc}$. We assume also that each swap introduces depolarizing noise parametrized by a swap quality $s_q$ and that memory decoherence is described by a $T_1$, $T_2$ process. The parametrization of depolarizing noise we consider is clarified in Appendix \ref{sec:am_validation}, and $T_1$, $T_2$ noise processes are discussed in more detail in the same Appendix. We further assume that entanglement swapping, although noisy, is deterministic. We note that this model is quite abstract. It could, in principle, describe the behaviour of any repeater of the processing node type, examples being NV centers and trapped ions. By this we mean that it is possible to map the parameters in a physically accurate model of an NV center or trapped ion to this smaller set of more abstract parameters. In fact, we did exactly this for NV centers in order to validate this model, as laid out in Appendix~\ref{sec:am_validation}. It is important to note that in this mapping we considered induced dephasing noise instead of the usual memory dephasing. Furthermore, atomic ensemble based repeaters could be described by considering non-deterministic entanglement swaps and enriching the model with a swap success probability parameter, but this lies beyond the scope of this work.
\subsection{Genetic Algorithms}

Evolutionary Algorithms (EAs) have been shown to have an advantage over conventional gradient-based methods in finding global minima in multimodal functions whose search space is not well known \cite{vikhar2016evolutionary}, although we stress that this is not guaranteed. They are also robust to noise in data and easy to parallelize. There are multiple approaches within the umbrella of EAs, with prominent examples being genetic algorithms (GAs) \cite{holland1992adaptation}, evolution strategy \cite{beyer2002evolution}, differential evolution \cite{storn1997differential} and particle swarm optimization \cite{kennedy1995particle, shi1998modified}. In this work we have used GAs, a search heuristic inspired by the theory of evolution. We  limit ourselves to a high-level overview of GAs. For a comprehensive introduction, we direct the interested reader to \cite{goldberg2006genetic}. 

We start with a population of randomly generated individuals. In our case, each individual in a population is a set of values for the parameters of the abstract model introduced in section~\ref{sec:abstractmodel}. The GA generates new individuals in an iterative process, with each iteration being known as a generation. In each generation, the cost function is evaluated for every member of the population, the resulting value being known as the fitness. A subset of the population is then selected according to a fitness-dependent rule, in which higher-fitness solutions are more likely to be chosen. New individuals are then generated through random crossover and mutation operations. The new population is used for the following iteration of the algorithm, meaning that the simulation is run with the new individuals (i.e. sets of abstract model parameters) as input and the cost function is computed using the simulation outputs. The algorithm can terminate after a set number of generations or once some predefined condition is attained. Exploration of the search space is assured by the crossover and mutation-driven recombination of solutions, whereas fitness-based selection ensures exploitation of minima.


GAs come in several different flavours. We defer to Appendix \ref{sec:appendiGA} for details on our particular implementation.

\subsection{\textit{smart-stopos}}
\label{sec:smart-stopos}
The simulation tools we use are computationally heavy and produce large amounts of data. In order to make good use of them and extract useful information from said data, we need a systematized way of feeding input parameters to the simulations in batches, run the simulations on a high-performance computing (HPC) cluster using \textit{stopos} \cite{stopos}, feed the outputs to the optimization algorithm and iterate this procedure. To these ends, we made use of \textit{smart-stopos} (freely available at~\cite{smartstopos}), a set of tools we developed to allow for parameter exploration and optimization, both locally and in an HPC setting. 
We used GAs in this work but in principle any other algorithm could be plugged in, provided that it can be run with only simulation inputs and outputs. Furthermore, we note that we used NetSquid but our methodology could also be made to work with any other quantum network simulator. For more details on the use of \textit{smart-stopos}, we direct the interested reader to Appendix~\ref{sec:appendixsmartstopos}.
\subsection{Process Overview}
\label{sec:processoverview}
\begin{figure}[!ht]
\centering
\includegraphics[width=0.5\columnwidth]{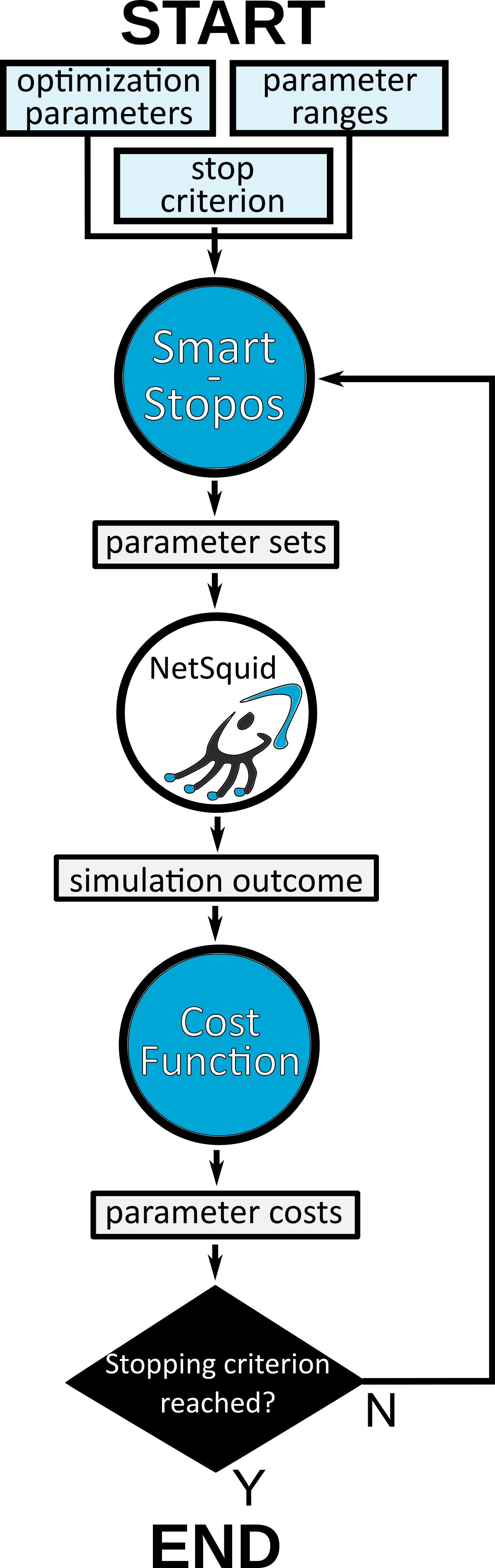}
\caption{Overview of our optimization process. The user inputs the desired optimization parameters, their ranges and a stopping criterion.  \textit{smart-stopos} generates sets of parameters in the allowed range and feeds them to the NetSquid simulation. The outputs of the simulation are used to compute the cost associated to each parameter set, which in turn is used by \textit{smart-stopos} to generate new parameter sets. This process is repeated until the stopping criterion is reached. In our particular case, the optimization parameters are the parameters defining the abstract repeater model introduced in~\ref{sec:abstractmodel}, the relevant simulation outputs are the fidelity and generation rate of end-to-end entangled states and the cost function is the one defined in~\ref{sec:cost}.}
\label{fig:diagram_optimization_process}
\end{figure}

We will now show how the tools we introduced can be pieced together to answer the question of what the minimum requirements are on the quality of quantum repeaters. To that end, we show in Figure~\ref{fig:diagram_optimization_process} a diagram of the workflow of our methodology.

The process is started by defining the parameters to be optimized and their allowed range of values. This information, together with a termination criterion, is passed to \textit{smart-stopos}, which then randomly generates sets of parameters within the defined ranges. Each of these sets of parameters is fed to the NetSquid simulation, which outputs an end-to-end entangled state and the time its generation took, allowing us to compute the fidelity with the ideal Bell state and the entanglement generation rate. These metrics, together with the parameter values and the baseline values, are used to compute the cost function, as defined in Equation~\ref{eq:costfunction}. This process is then repeated for each set of parameters. The ensemble of parameter sets and respective costs are given as input to \textit{smart-stopos}, which generates new sets of parameters using our GA. The process repeats until the termination criterion is reached. The final output is the minimum value of the cost function found by the algorithm, which in this case corresponds to an answer to the question of what are the minimum requirements on a quantum repeater.

Figure~\ref{fig:diagram_optimization_process} makes the modularity of our approach clear.
Any of the building blocks of our process, namely the optimization algorithm used by \textit{smart-stopos}, NetSquid simulation and cost function, can be swapped out without changes to the overall workflow. For example, if we wanted to apply our methodology to a simulation of a repeater chain of trapped ions, all we would have to do would be to replace our abstract model NetSquid simulation for an appropriate trapped ions simulation. Similarly, to answer a different optimization question one just has to redefine the cost function.

\subsection{Challenges in Applying GAs to Quantum Systems}
We came across some challenges when applying GAs to simulations of quantum systems. Some of these were of a practical nature, and others were more fundamental. We will now give an overview of what these issues were, and how we overcame them.

\subsubsection{Practical Challenges}
\begin{figure}[!ht]
\centering
\includegraphics[width=\columnwidth]{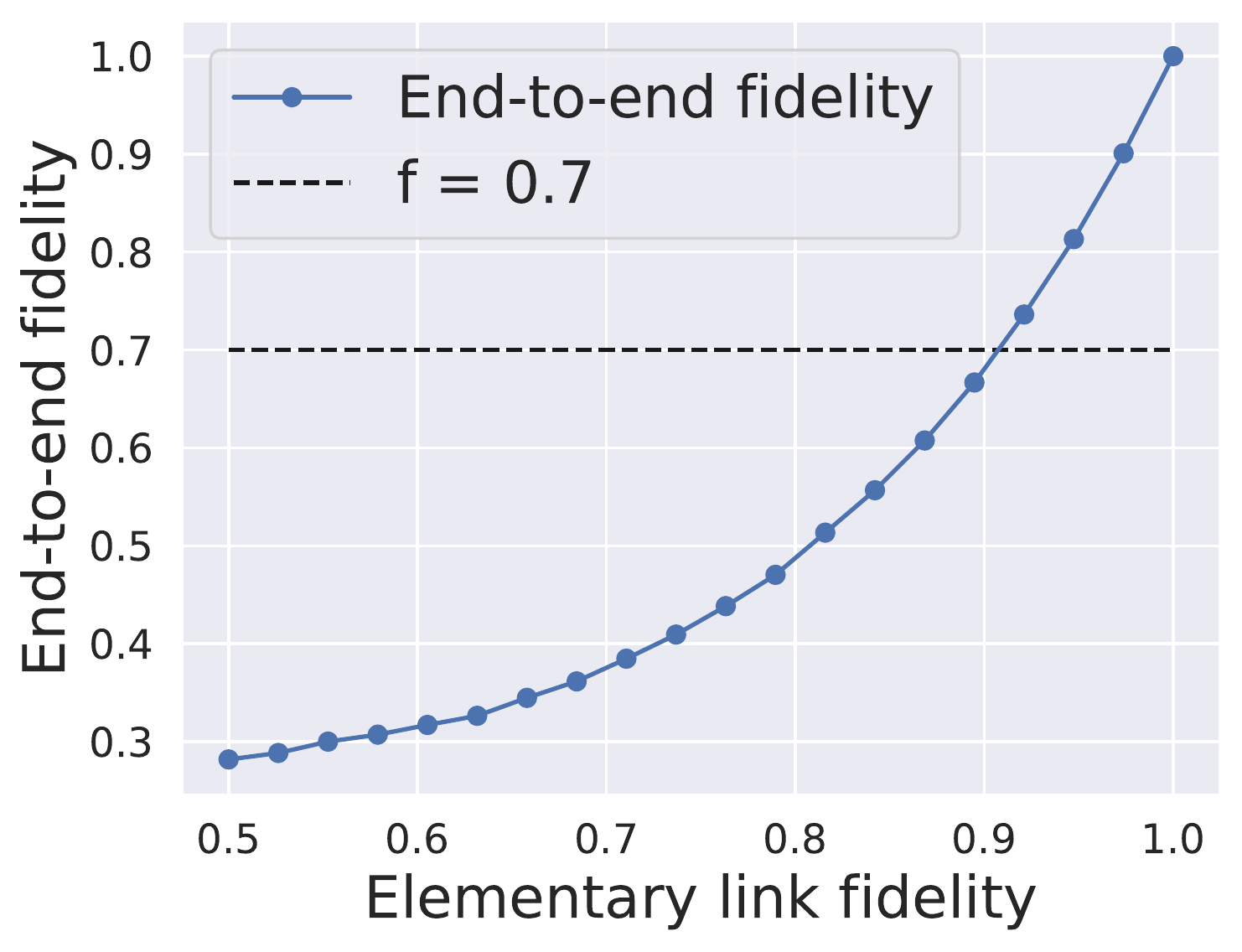}
\caption{Variation of end-to-end fidelity across five equally spaced nodes as the elementary link fidelity is varied and the other four parameters are kept at their perfect values. The value of the elementary link fidelity that results in an end-to-end fidelity of $0.7$ is at the intersection of the two lines in the plot, being just above $0.9$ in this case.}
\label{fig:sensitivityanalysis_elf}
\end{figure}

We came across two practical challenges: (i) the size of the parameter space and (ii) the amount of data generated. (i) is due to the complexity of quantum repeater modelling. In general the search space may be big, but in our illustrative example of the abstract model introduced in~\ref{sec:abstractmodel} it is manageable. We nevertheless introduced a pre-processing procedure for restricting the parameter space, as we believe it would be useful when considering use cases with larger parameter spaces. This procedure consists of performing sensitivity analysis for each of the five parameters individually, i.e. holding four parameters constant and running simulations varying the fifth one from its baseline value to its perfect one. As an example of how this can reduce the search space, we show in Figure~\ref{fig:sensitivityanalysis_elf} the variation of the end-to-end fidelity with the elementary link fidelity when all other parameters are kept at their perfect values. The optimal set of parameters for this setup will certainly contain less-than-perfect values, so the elementary link fidelity of this set will be higher than the one found using this sensitivity analysis, so we can safely restrict the search space for this parameter in GA optimizations runs to the interval $[f_{\text{perf}}, 1.]$, where $f_{\text{perf}}$ is the elementary link fidelity that results in an end-to-end fidelity of $0.7$ when all other parameters are perfect.

Another practical challenge is the sheer amount of data that is produced. For each setup we consider we run our simulations for hundreds of different sets of parameters at each optimization step, with each set of parameters being in turn run a hundred times. In order to systematically and efficiently process all of this data, we developed \textit{smart-stopos}, as detailed in Section~\ref{sec:smart-stopos}.
\subsubsection{Fundamental Challenges}
Fundamental challenges occur due to the fact that quantum systems produce inherently non-deterministic outputs. This can be problematic if the cost function has terms that are step functions, which is our case. For a concrete example, let us say that in generation $34$ of the optimization procedure, the GA found a set of parameters that result in an entanglement generation rate of $1.05$ Hz, just above the desired threshold. In generation $35$, this parameter set would again be fed into the simulation. However, this time around, due to statistical fluctuations, the simulation outputs an entanglement generation rate of $0.99$ Hz, just below the threshold. Since the cost function defined in Equation~\eqref{eq:totalcost} assigns a very high cost to any solution that does not attain the performance metrics, this solution would in generation $35$ have a very high cost function value. This means that it would almost certainly not be chosen as a parent for the following generation, and the algorithm would effectively lose it. This is a problem, as it results in the algorithm losing a good solution and potentially wasting computation time finding it again. 

There are several possible solutions to this problem. The one we chose, due to its simplicity, was to run the simulation multiple times for each set of parameters and compute the value of the cost function using the average end-to-end fidelity and entanglement generation rates. Running the simulations multiple times provides some security against statistical fluctuations, although it increases the computation time. We found empirically that running the simulation $100$ times for each set of parameters represents a good trade-off between minimizing fluctuations and keeping computation times feasible. 

Another possible solution that we also explored was to use a smoother function, such as a sigmoid, instead of a sharp step function. This would in principle address the problem we mentioned of a set of parameters being heavily penalized because its metrics dipped just below the targets due to statistical fluctuations. For a smoother function, such fluctuations would lead to small fluctuations in the value of the cost function. There are however some issues with this solution. Since the function is smoother, it no longer acts as a hard constraint, which is the behaviour we are looking for. What we mean by this is that a solution whose performance metrics are slightly below the targets will only be lightly penalized. It might thus have a lower cost function value than a solution with better, i.e. more expensive, parameters that attains the performance metrics. In less technical terms, this translates as the cost function not being well aligned with the stated optimization goal. 

This concludes the introduction of the optimization methodology we propose. The rest of the paper concerns itself with two questions: (i) is our methodology valid, addressed in Section~\ref{sec:validation} and (ii) what results do we get when we apply it, addressed in Section~\ref{sec:usecase}.

\section{Validation} \label{sec:validation}
As we stated in the previous section, before we apply our methodology we must validate it. By this we mean that we must verify that the methodology we propose for applying genetic algorithms to simulations of quantum networks can produce meaningful results. We can see this validation as being split into two different steps. One, benchmarking the genetic algorithms i.e., evaluating how well they perform and two, validating that the methodology is sound. The first step will be accomplished by applying our specific implementation of genetic algorithms to the optimization of common benchmarking functions and comparing their performance to that of implementations found in the literature. The second step will consist of applying our methodology to a chain of evenly-spaced nodes generating Werner states, for which analytical expressions for the end-to-end fidelity and entanglement generation rate in terms of repeater parameters can be found. Having these expressions, we can compute what are the repeater parameters that minimize the cost function. If our GA approach is capable of finding this solution, we have compelling evidence that our methodology would also perform well when applied to the more realistic cases we are interested in, for which analytical results cannot be readily derived.

We have also validated the abstract model we use in our simulations against a more physically accurate model of NV center-based repeaters. These results are shown in Appendix~\ref{sec:am_validation}. 
\subsection{Benchmarking Genetic Algorithms}
\begin{figure}[!ht]
  \centering
  \begin{tabular}[b]{c}
    \includegraphics[width=0.6\linewidth]{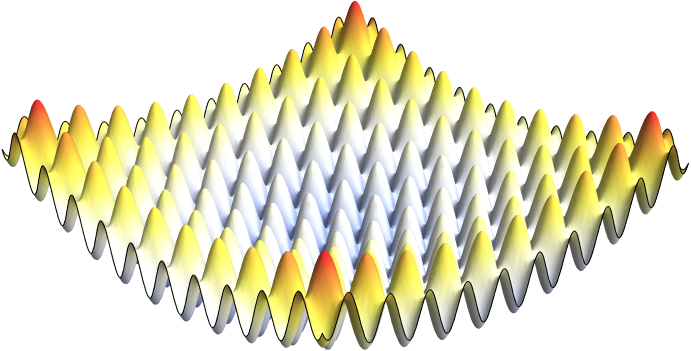} \\
    \small (a) Rastrigin's function.
  \end{tabular} \qquad
  \begin{tabular}[b]{c}
    \includegraphics[width=0.6\linewidth]{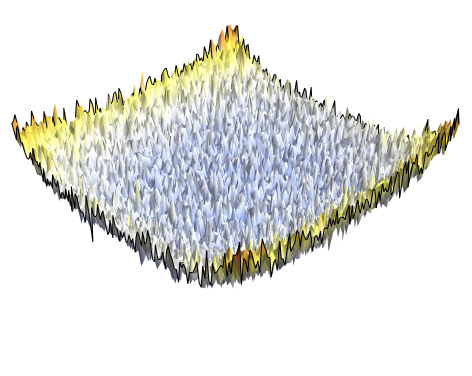} \\
    \small (b) Quartic function.
  \end{tabular}
  \caption{Plot of the 2-dimensional versions of (a) Rastrigin's function and (b) quartic function. The multiple minima of Rastrigin's function and the noisy landscape of the quartic function can be clearly seen.}
  \label{fig:benchmarking_functions}
\end{figure}

In order to evaluate the performance of GAs and how it is affected by the algorithm's hyperparameters, several benchmarking functions have been defined~\cite{digalakis2001benchmarking}. These are designed to test how well each GA implementation handles cost functions with given properties. For example, if we expect the function we want to optimize to be noisy, i.e. to have the output for a given input randomly oscillate each time the function is called, we should benchmark the GA against a noisy function, such as the quartic function, defined in Equation~\eqref{eq:quartic}

\begin{equation}
    f_q(\mathbf{x}) = \sum_{k=1}^{30} \left(kx_k^4 + \mathcal{N}(0, 1)\right) \quad -1.28 \leq x_k \leq 1.28,
\label{eq:quartic}
\end{equation}
where $\mathcal{N}(0,1)$ is a normal distribution with mean $0$ and standard deviation $1$.
This function, plotted in the bottom half of Figure~\ref{fig:benchmarking_functions}, is a unimodal function padded with Gaussian noise. Therefore, a GA that performs poorly on it will also perform poorly on any function with noisy outputs.

\begin{figure}[ht]
  \includegraphics[width=.5\columnwidth]
    {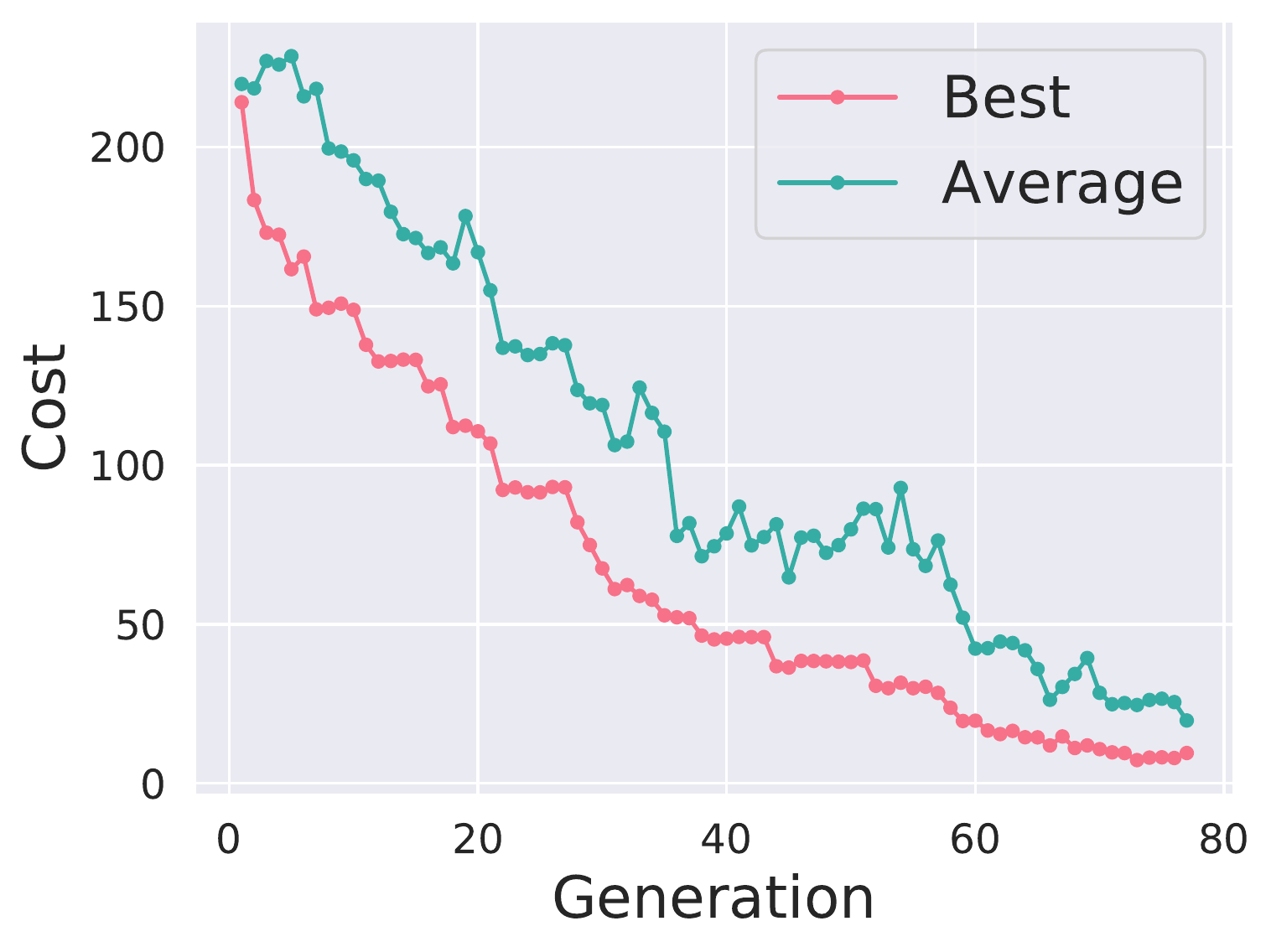}\hfill
  \includegraphics[width=.5\columnwidth]
    {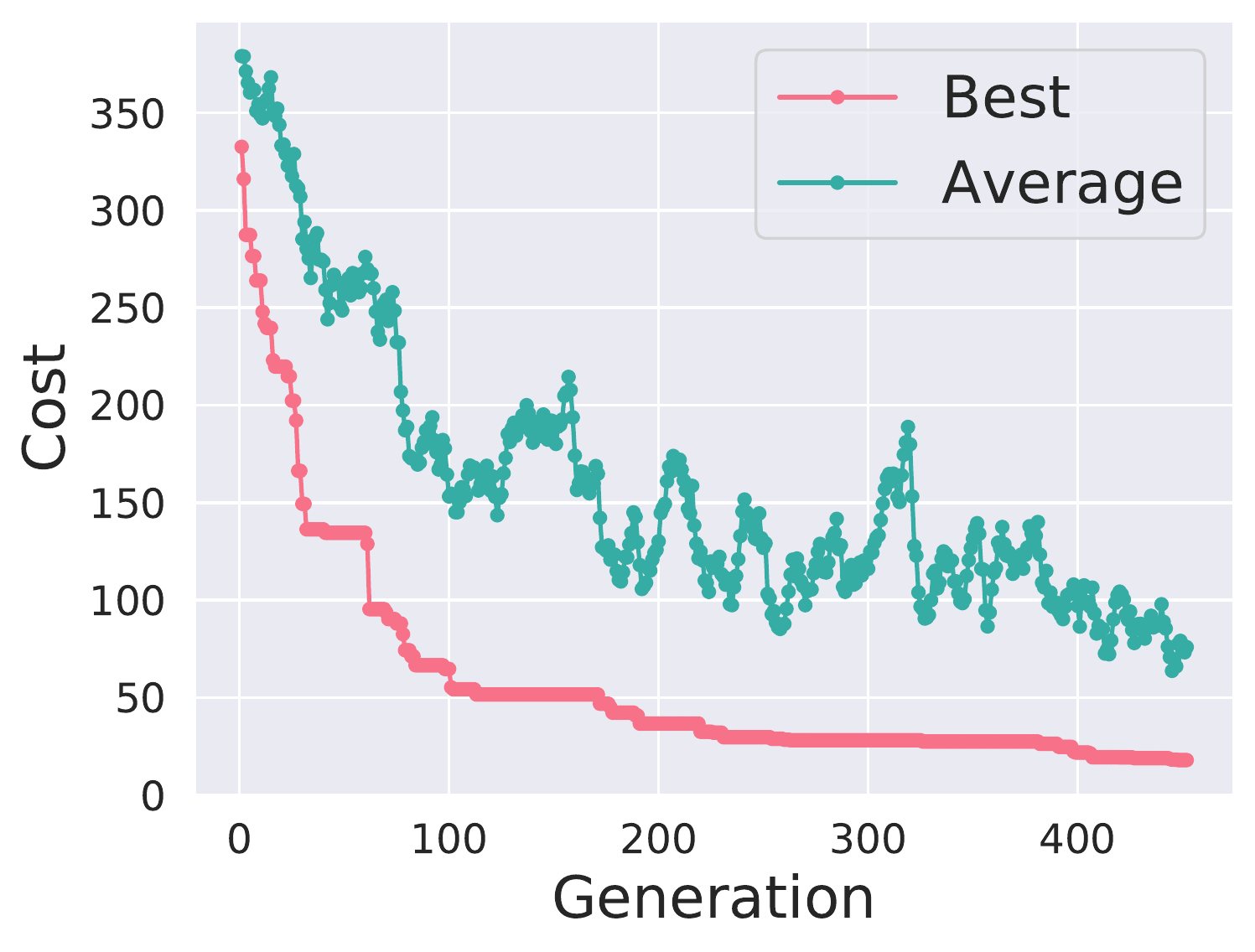}

  \caption{Evolution of the cost of best solution (red) and population average (green) for the quartic function (left) and Rastrigin's function (right) over 75 and 400 generations, respectively. All approach zero, the global minimum of both cost functions, with the average cost being consistently higher than the best cost, as expected. This indicates that our GA implementation is capable of finding good solutions for said functions.}
  \label{fig:costs_benchmarking}
\end{figure}

Taking this into account, we chose two functions to benchmark our GA implementations. This choice was made by taking into account which of the functions best represented the cost landscape we expect our problem to have. Since the quantum nature of our simulations implies that they will necessarily be noisy in the above-defined sense, we will choose the quartic function as a benchmarking function. Furthermore, we expect that the landscape of the cost function defined in Equation~\eqref{eq:costfunction} will have multiple local minima, corresponding to different sets of parameters that satisfy the imposed constraints on end-to-end fidelity and entanglement generation rate. With this in mind, we also chose Rastrigin's function, defined in Equation~\eqref{eq:rastrigin}.

\begin{equation}
\footnotesize
    f_r(\mathbf{x}) = 200 + \sum_{i=1}^{20} \left(x_i^2 - 10 \cos\left(2\pi x_i \right)\right), -5.12 \\\leq x_i \leq 5.12
\label{eq:rastrigin}
\end{equation}
For illustrative purposes, the 2-dimensional version of Rastrigin's function is shown on the top half of Figure~\ref{fig:benchmarking_functions}. It can be seen that it has a very bumpy landscape, with a global minimum at $\textbf{0}$, in the center of the plotted region. Its many local minima render it a challenging benchmark for GAs.
We applied our GA implementation to both of these functions, with the results being plotted in Figure~\ref{fig:costs_benchmarking}. The hyperparameters used for these optimization runs were chosen according to the guidelines given in \cite{digalakis2001benchmarking} and population selection was done using the Roulette Wheel method \cite{10.5555/534133}. For an explanation of the Roulette Wheel method we point the interested reader to Appendix \ref{sec:appendiGA}. We see that, for both functions, the cost averaged over the whole population and the cost of the best member at each generation approach their global minimum, 0. Furthermore, the performance of our implementation is in line with that of those in \cite{digalakis2001benchmarking}, which indicates that our GA is capable of handling both noisy and multimodal functions.
We note that we could, by further tuning some of the algorithm's hyperparameters, obtain a marginally better performance on these benchmarking functions. However, since our goal is only to verify that our implementation is correct and performs reasonably well for the type of cost landscapes that we expect to encounter, we abstain from doing so.
\subsection{Validating on Werner Chains}
The previous section focused on benchmarking the performance of the GA, but the question of whether applying GAs to repeater chain optimization problems can produce good results remains. In order to answer it, we consider the simple scenario of a chain of $3$ nodes generating Werner states, and we pose the question of what are the worst parameters that can deliver an end-to-end entangled pair of fidelity $0.6$ every second. Similarly to the abstract model presented in earlier sections, the nodes in the chain generate elementary links of fidelity $F_{EL}$ with success probability $p_{suc}$ and depolarizing noise parametrized by $s_q$ is applied after entanglement swaps. This is a problem for which we can analytically find expressions for the end-to-end rate and fidelity, and thus for the ideal value of the cost function. We expect that the structure of this problem is similar to that of the one we want to tackle. By this we mean that we expect its cost landscape to show some of the same features as our target problem, namely multiple minima and noisiness. Therefore, despite being simpler, good performance in this problem should indicate that our approach is valid. For details of how we derived analytical results for this setup we defer the interested reader to Appendix~\ref{sec:appendixwerner}.

\begin{figure}[!htpb]
\centering
\includegraphics[width=\columnwidth]{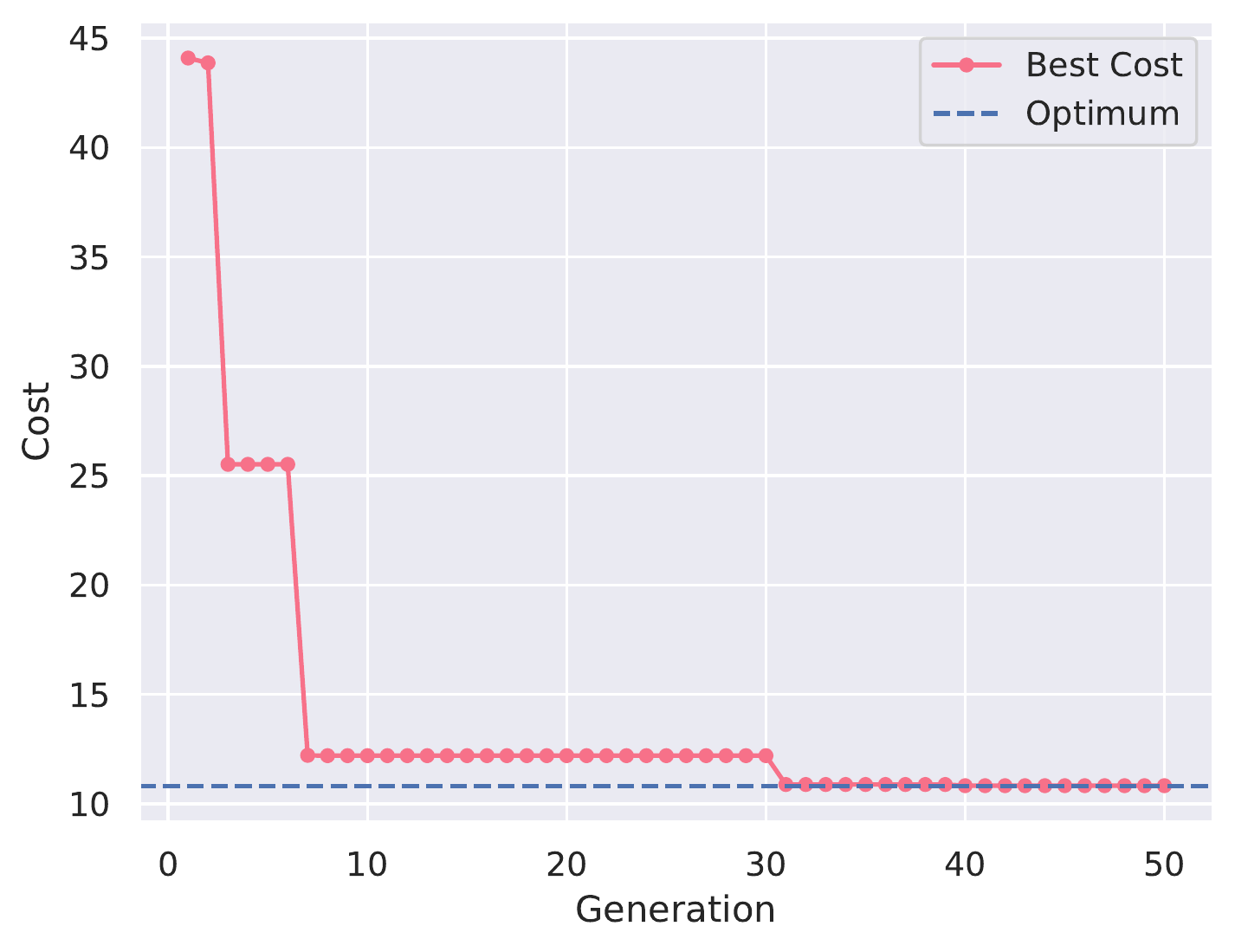}
\caption{Evolution of the lowest value of the cost function over 50 generations. After little more than 30 generations the algorithm finds the parameter set that optimizes the cost function. This optimum is marked in the figure by a blue dashed line.}
\label{fig:cost_vs_generation_werner}
\end{figure}

In Figure~\ref{fig:cost_vs_generation_werner}, we show the evolution of the cost of the best individual in each generation obtained by applying the GA-based method to the setup we described. Also present in the plot, in a dashed line, is the optimum cost.

The cost function drops to the global optimum at around the 30 generation mark, indicating that the algorithm is capable of finding the worst set of repeater parameters satisfying the benchmarks we set. This is then a good indicator that our methodology is well-suited to the optimization of entanglement generation in repeater chains.
\section{Evaluation: Use Cases} \label{sec:usecase}
Having validated our methodology, we applied it to two use cases demonstrating its power and potential usefulness. In the past decade, NV centers have been demonstrated to be capable of generating remote entanglement between matter memories with long coherence times~\cite{hensen2015loophole, humphreys2018deterministic, bradley2019ten}, establishing them as promising candidates for the realization of scalable quantum repeaters~\cite{rozpkedek2019near}. A better understanding of hardware requirements would then be useful in illuminating the path towards scalable NV-based quantum repeaters. We thus used the abstract model of NV-type states that we introduced in Section~\ref{sec:abstractmodel} in the simulations of all use cases. We furthermore chose to consider, for simplicity, SWAP-ASAP protocols with no memory cut-offs (see section~\ref{sec:abstractmodel} for an explanation).

\begin{figure}[!htpb]
\centering
\includegraphics[width=0.7\columnwidth]{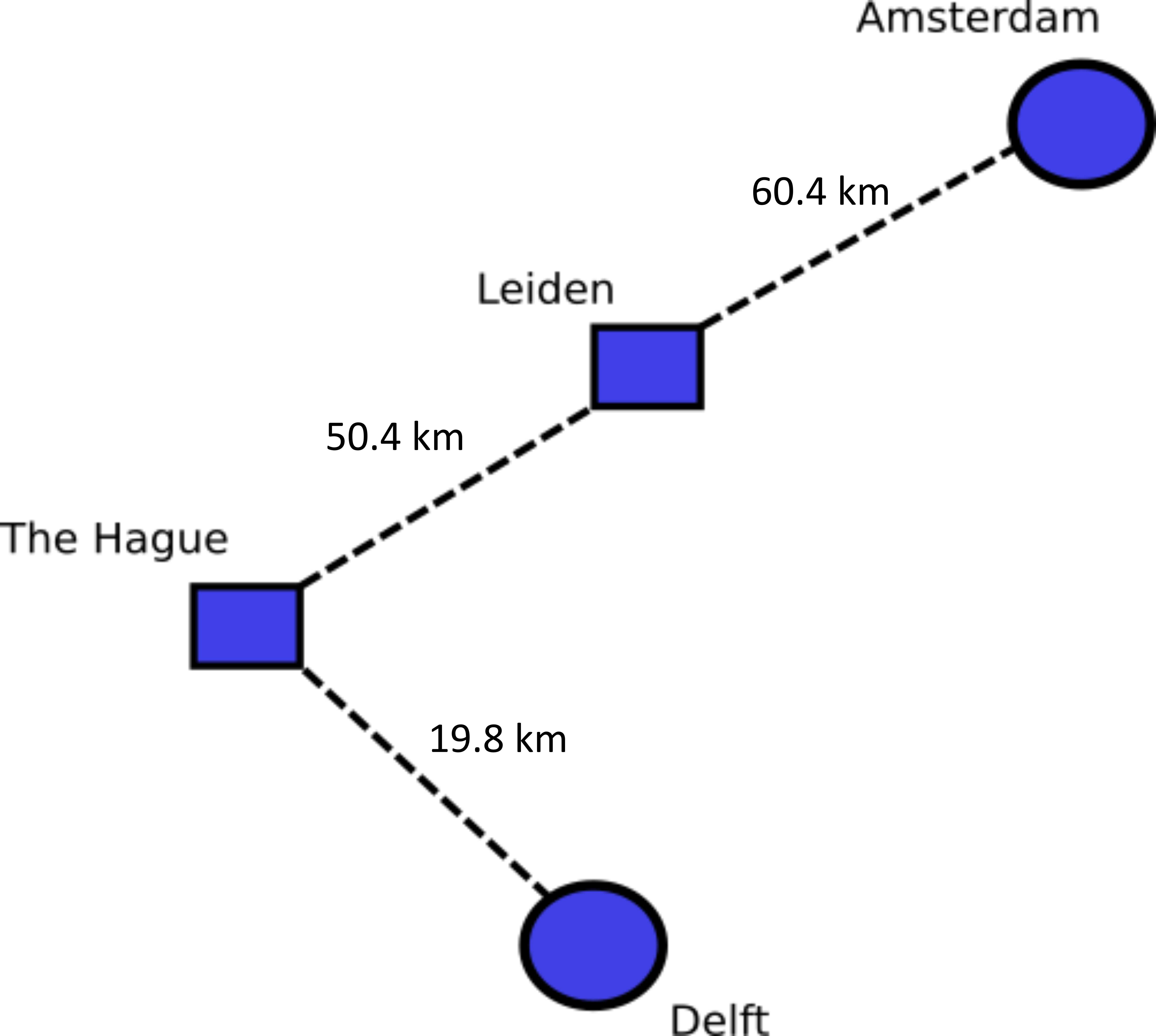}
\caption{Visualization of the quantum network we will consider. The end nodes, represented by circles, are placed in Delft and Amsterdam. The repeater nodes, represented by squares, are placed in The Hague and Leiden. The placement of the nodes roughly approximates their actual geographical location and the length of the fibers connecting them is included for reference.}
\label{fig:real_life_network}
\end{figure}

Another roadblock in the way of the quantum internet is that even when quantum repeater technology is at deployment stage, it is expected that it will be very costly. One way of rendering the implementation of quantum networks more cost-effective is to take advantage of preexisting infrastructure by using previously deployed optical fiber networks~\cite{rabbie2020designing}. With this in mind, we used real-life fiber data of the Netherlands. This was made available to us by SURF, a network provider for Dutch education and research institutions. We considered a repeater chain with nodes in Delft, The Hague, Leiden and Amsterdam, as depicted in Figure~\ref{fig:real_life_network}, as this is an example of a possible near-term quantum network in the Netherlands. We use real fiber length and attenuation in our simulations. We chose Delft and Amsterdam as the end nodes of the chain as out of these four cities they are the most distant pair. The baseline values used for computing the value of the cost function for each set of parameters were obtained from actual state-of-the-art experimental results using NV centers. The process through which we converted these experimental results to our abstract model parameters is described in detail in Appendix~\ref{sec:appendixbaseline}. We set as performance targets end-to-end fidelity $F_{min} = 0.7$ and end-to-end entanglement generation rate $R_{min} = 1$ Hz. The value of $F_{min}$ was chosen to ensure that we remain in the regime where the agreement between the abstract model and the detailed NV model is good.

In order to study the effects of internode distance, chain length and number of repeaters we further applied our methodology to chains of equally spaced nodes with varying numbers of repeaters. In one case, we kept the internode distance fixed, and in the other we kept the total length fixed as we varied the number of repeaters. More concretely, we considered (i) a chain of equally spaced nodes spanning $800$ km and (ii) a chain with an internode distance of $100$ km. For each of these, we considered the cases of $3$, $5$, $10$ and $12$ repeater nodes. The baseline parameter values are computed in the same manner as in the previous use case, so we again defer to Appendix~\ref{sec:appendixbaseline} for details. We also consider the same target performance metrics as in the previous use case.
\subsection{Results}
\subsubsection{Real Network}\label{sec:real_network}
We will now show the main results obtained by applying our methodology to the network introduced in Figure~\ref{fig:real_life_network}. 

\begin{figure}[!htpb]
\centering
\includegraphics[width=\columnwidth]{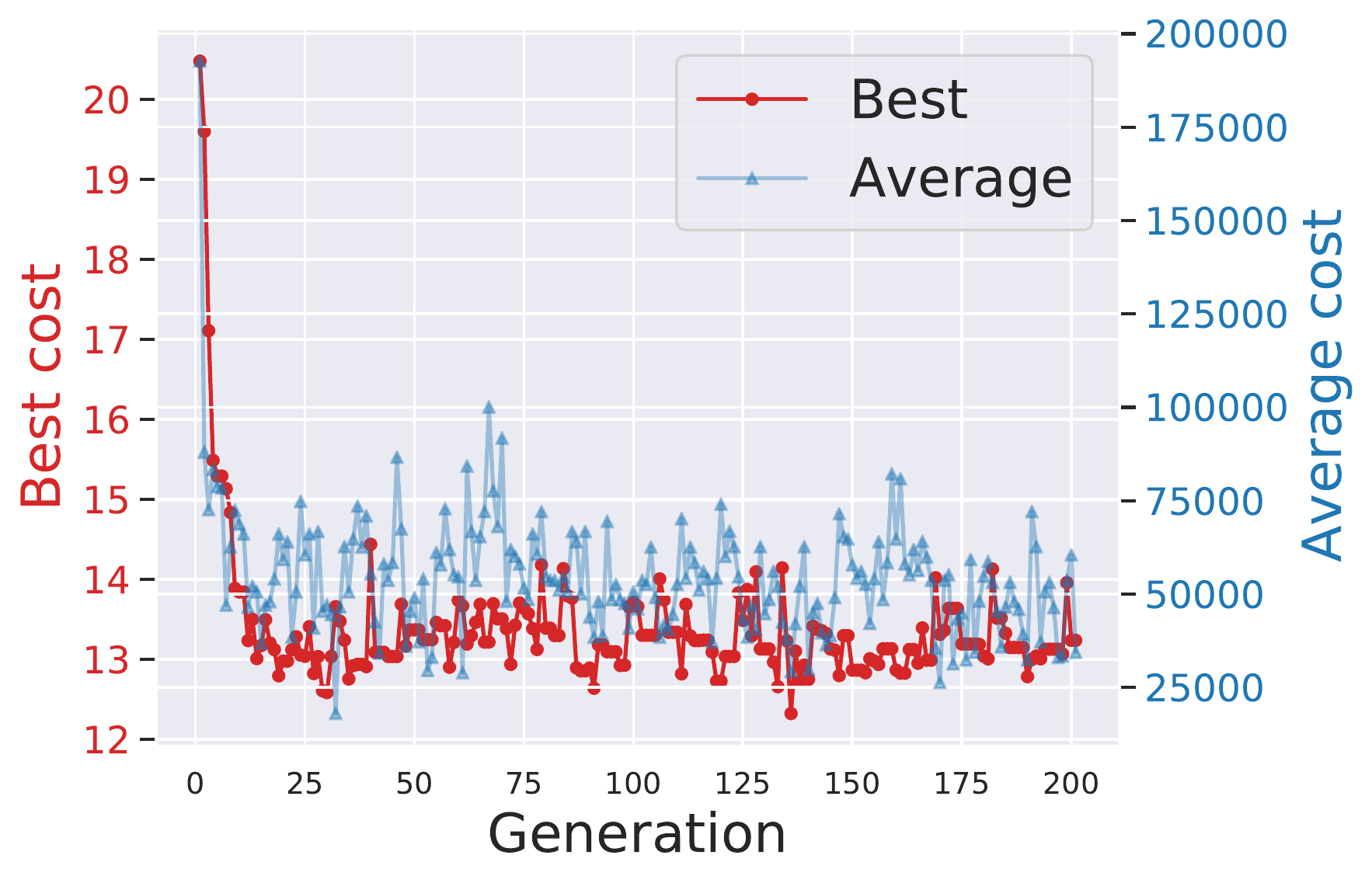}
\caption{Evolution of the best (red circles) and average (blue triangles) values of the total cost function for the use case discussed in Section~\ref{sec:real_network}. The best value converges to roughly $13$, whereas the average oscillates around $50000$ (note the differing scales). The high values of the average cost throughout the optimization process are due to the mutation process, which sometimes produces solutions that do not fulfill the target metrics. Our simulation was run $100$ times for each individual.}
\label{fig:cost_vs_generation_real}
\end{figure}

In Figure~\ref{fig:cost_vs_generation_real} we show the best and average values of the total cost function (Equation~\eqref{eq:totalcost}) as a function of the optimization step. Contrasting with~Figure~\ref{fig:costs_benchmarking}, we see that (i) the average value of the cost function remains significantly higher than the best value and that (ii) the best value per generation oscillates. The first observation is explained by the combination of the inherent randomness of the GA and the fact that we used step functions for the cost. A GA generates new candidate solutions through a process of mutation and recombination, as detailed in Appendix~\ref{sec:appendiGA}. While these processes allow for a thorough exploration of the parameter space, they may also produce solutions that fall outside the defined target metrics. The step functions in the cost ensure that such solutions will be heavily penalized, explaining the high average values of the cost function in Figure~\ref{fig:cost_vs_generation_real}. The second observation is also explained by a combination of two factors, namely the already mentioned step functions in the cost and the non-deterministic nature of our simulations. Since across different simulations for the same set of parameters there are fluctuations in the values of the end-to-end metrics, it might happen that these sometimes dip below the predefined targets. Due to the step function, the cost associated to this particular set of parameters will become much higher, meaning that it will no longer be the best solution. This effect can be minimized by running our simulations multiple times for each set of parameters, as discussed in Section~\ref{sec:processoverview}.

In Table~\ref{tab:solution_real_network} we show the parameters of the best solution found using our methodology. For comparison purposes, we also show the baseline values we considered.
\begin{table}[!ht]
\begin{tabular}{|c|c|c|c|c|c|}
\hline
         & $F_{EL}$ & $p_{suc}$ & $s_q$  & $T_1$   & $T_2$    \\ \hline
Baseline & 0.9698   & 0.004600  & 0.8590 & 10 h    & 4.9 ms   \\ \hline
Solution & 0.9806   & 0.09770   & 0.9414 & 10.23 h & 22.79 ms \\ \hline
\end{tabular}
\caption{Experimentally-derived baseline parameter values and values of the best solution found using our methodology for the use case discussed in Section~\ref{sec:real_network}. The biggest relative increases happen for $T_2$ and $p_{suc}$, suggesting that improving these parameters is key for achieving scalable NV-based repeaters.}
\label{tab:solution_real_network}
\end{table}
The biggest relative increases are in $p_{suc}$ and $T_2$, suggesting that induced dephasing noise is the biggest hurdle in the way of NV-based repeater technology. On the other end of the spectrum, the solution's $T_1$ value is barely higher than that of the baseline, indicating that $T_1$ coherence times in NV centers are already long enough. 
\subsubsection{Equally Spaced Nodes}\label{sec:eq_spaced_nodes}
We now show the main results obtained by applying our methodology to repeater chains of equally spaced nodes with different numbers of repeaters. To study how the overall length of a chain and the internode distance affect the solutions found, we considered two cases: (a) fixed chain length (FCL) and (b) fixed internode distance (FID). For both FCL and FID we applied our methodology to chains of $3$, $5$, $10$ and $12$ repeater nodes. We note that each data point in the plots shown in this section corresponds to the best solution found after 200 generations, with 150 population individuals per generation and 100 simulation runs per individual. Running our optimization procedure once with these parameters takes roughly 46 hours locally using a standard laptop, underlining the need for access to HPC centers. In fact, by using such a center, the computation time can be reduced to 2 hours, in the best case scenario of not having to wait for computing node availability. We note that vast majority of this is taken by quantum repeater simulations, with the time needed by the GA being negligible in comparison.

\begin{figure}[!htpb]
\centering
\includegraphics[width=\columnwidth]{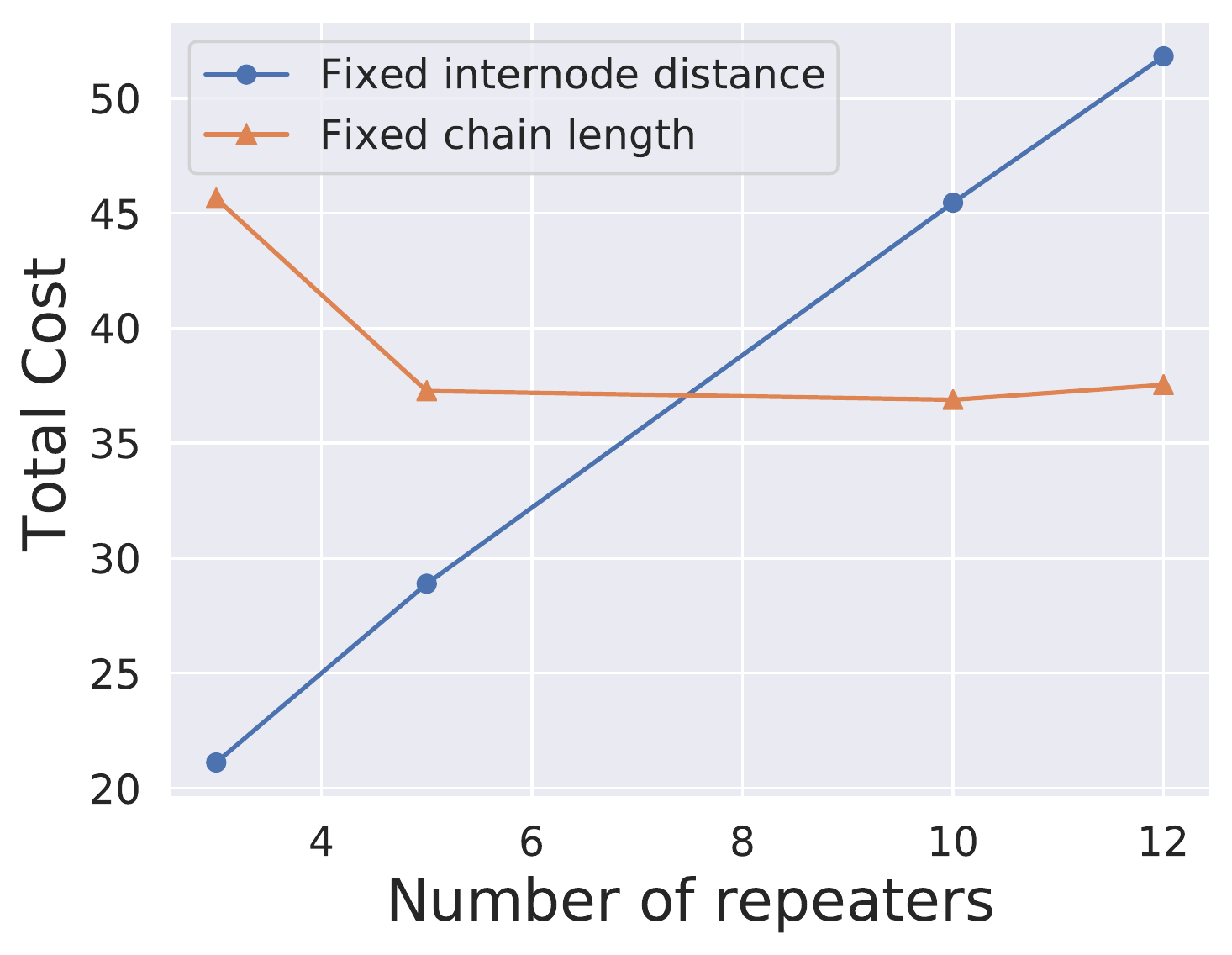}
\caption{Total cost, as defined in Equation~\ref{eq:totalcost}, of the best solutions found by our GA for setups with varying number of repeaters. The cost grows linearly for FID. There is no discernible pattern for FCL. Each data point corresponds to the best solution found after 200 generations, with 150 population individuals per generation and 100 simulation runs per individual.}
\label{fig:cost_vs_repeaters}
\end{figure}

In Figure~\ref{fig:cost_vs_repeaters} we show how the total cost of the best solution found varies with the number of repeaters in both cases. We observe a linear growth of the FID cost with the number of repeaters, which is not surprising: fixing the internode distance but increasing the number of repeater nodes corresponds to increasing the total length covered. In fact, the leftmost data point in Figure~\ref{fig:cost_vs_repeaters} corresponds to a chain spanning $400$ km, whereas the rightmost is associated to a chain spanning $1300$ km. We would expect connecting end nodes that are further apart to be a greater challenge due to the exponential growth in photon losses, which necessitates repeater parameters of higher quality. This does not apply to the FCL use case. All the data points in the associated curve correspond to a repeater chain that spans $800$ km and we observe in Figure~\ref{fig:cost_vs_repeaters} that the cost is slightly higher for the $3$-repeater setup. It was not \textit{a priori} obvious that this would be the case. A smaller number of repeaters implies that the swap quality and fidelity of the elementary link do not need to be as good, as there will be fewer swaps and hence less fidelity loss. On the other hand the elementary links are longer than in a setup with many repeaters, so the associated baseline values are worse (see Appendix~\ref{sec:appendixbaseline} for details). Any improvement then requires a higher parameter cost, as per Equation~\ref{eq:costfunction}.  

\begin{figure}[!htpb]
\centering
\includegraphics[width=\columnwidth]{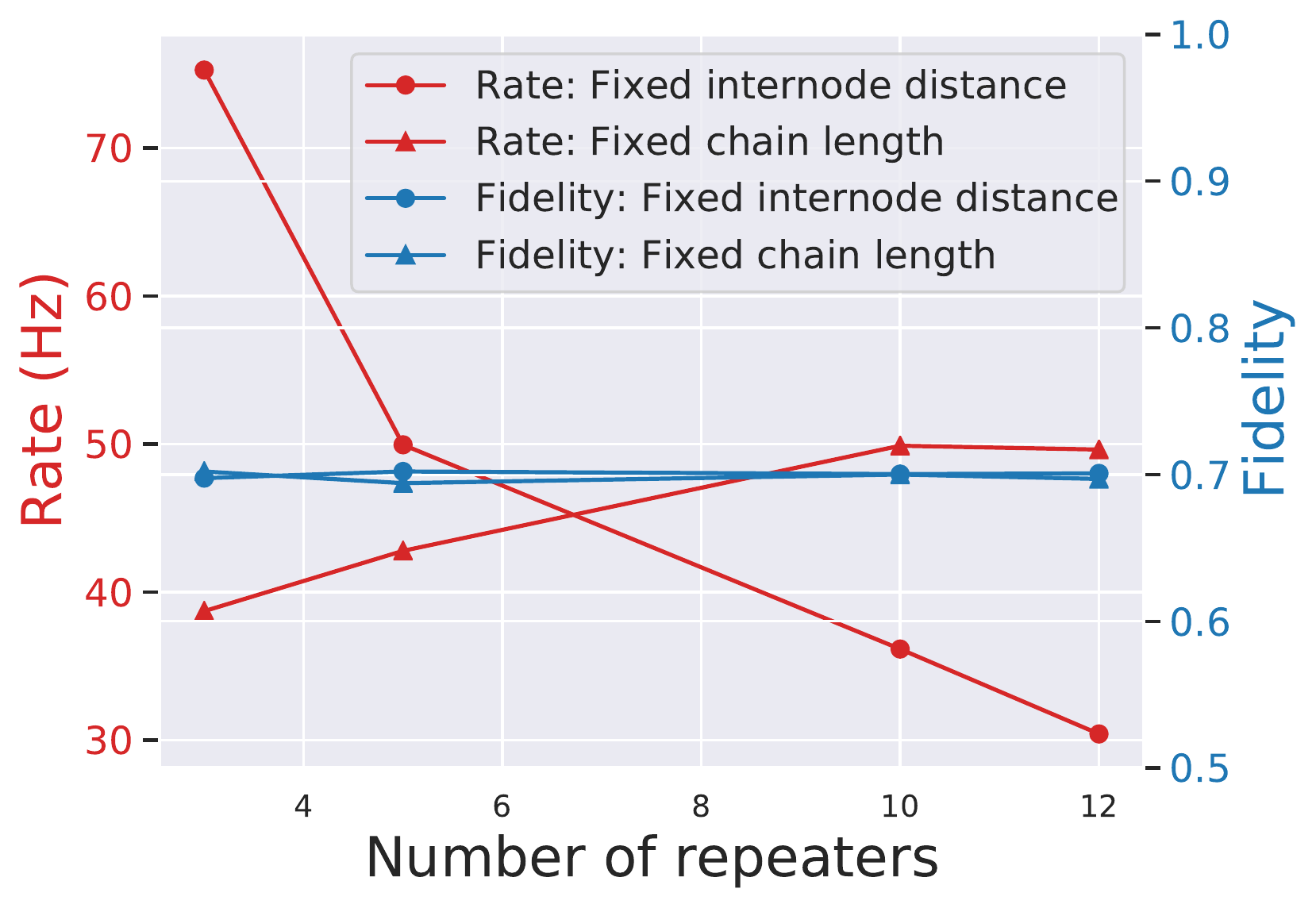}
\caption{Comparison of the metrics characterizing the best solutions found by our GA for each of the different setups. The end-to-end fidelity is very close to the goal of $0.7$ we defined, for both FID and FCL. On the other hand, the end-to-end entanglement rate is well above the $1$ Hz goal for both cases. For fixed internode distance, it decreases from roughly $80$ Hz in the $3$ repeater node setup to about $30$ Hz in the $12$ repeater node setup. For fixed chain length, it increases slightly from $40$ Hz in the $3$ repeater node setup to $50$ Hz in the $12$ repeaters setup. Each data point corresponds to $100$ runs of the simulation. The error bars are smaller than the markers.}
\label{fig:metrics_vs_repeaters}
\end{figure}

To further explore how the solutions found vary, we plot in Figure~\ref{fig:metrics_vs_repeaters} the end-to-end fidelity and entanglement generation rate of these solutions against the number of repeaters in the chain. We see that, for both use cases and all numbers of repeaters, the end-to-end fidelity is very close to $0.7$. On the other hand, the rate decreases from around $80$ Hz to $30$ Hz as the number of repeater nodes increases from $3$ to $12$ at FID and it increases slightly from $40$ Hz to $50$ Hz with the number of repeaters at FCL. While the fidelities obtained are what we expected, since the limit we imposed via the cost function was $0.7$, the same is not true for the rates. The penalty term we added to the cost function only comes into effect if the rate drops below $1$ Hz, so there is no benefit in terms of the cost to have a solution that results in a rate of e.g. $50$ Hz versus one of $1$ Hz. We would thus expect the best solutions to have rates close to $1$ Hz, which was not the case.
\begin{figure}
    \centering
  \includegraphics[width=\columnwidth]
    {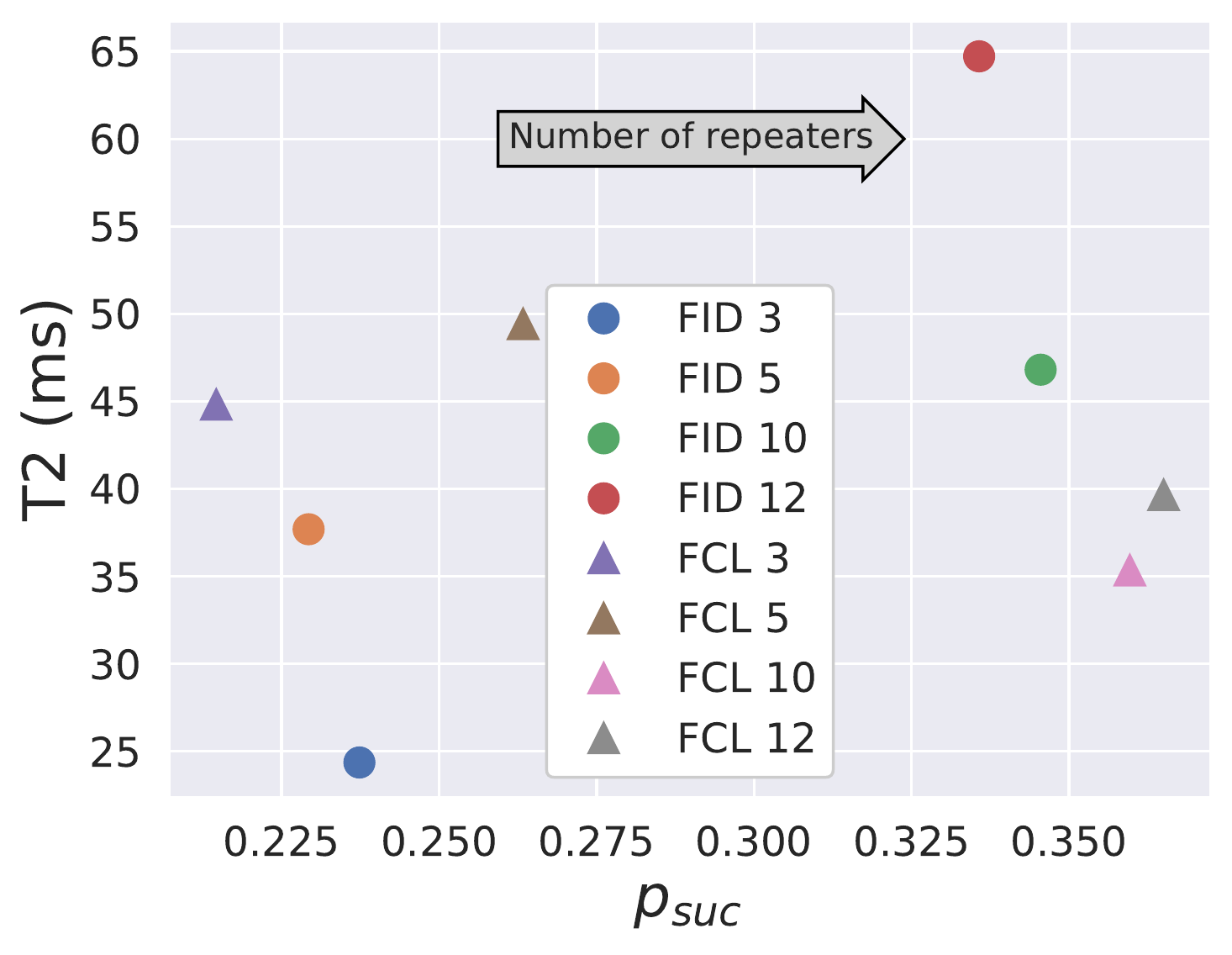}
    \caption{$T_2$ versus $p_{suc}$ of the best solution for each setup. More repeaters require higher $p_{suc}$. Furthermore, values of $T_2$ are higher for higher numbers of repeaters at FID, but lower at FCL. Each data point corresponds to the best solution found after 200 generations, with 150 population individuals per generation and 100 simulation runs per individual.}
    \label{fig:t2_psuc_best_solutions_both_setups}
\end{figure}\begin{figure*}[!ht]
\centering
\begin{tabular}[b]{c}
    \includegraphics[width=0.32\textwidth]{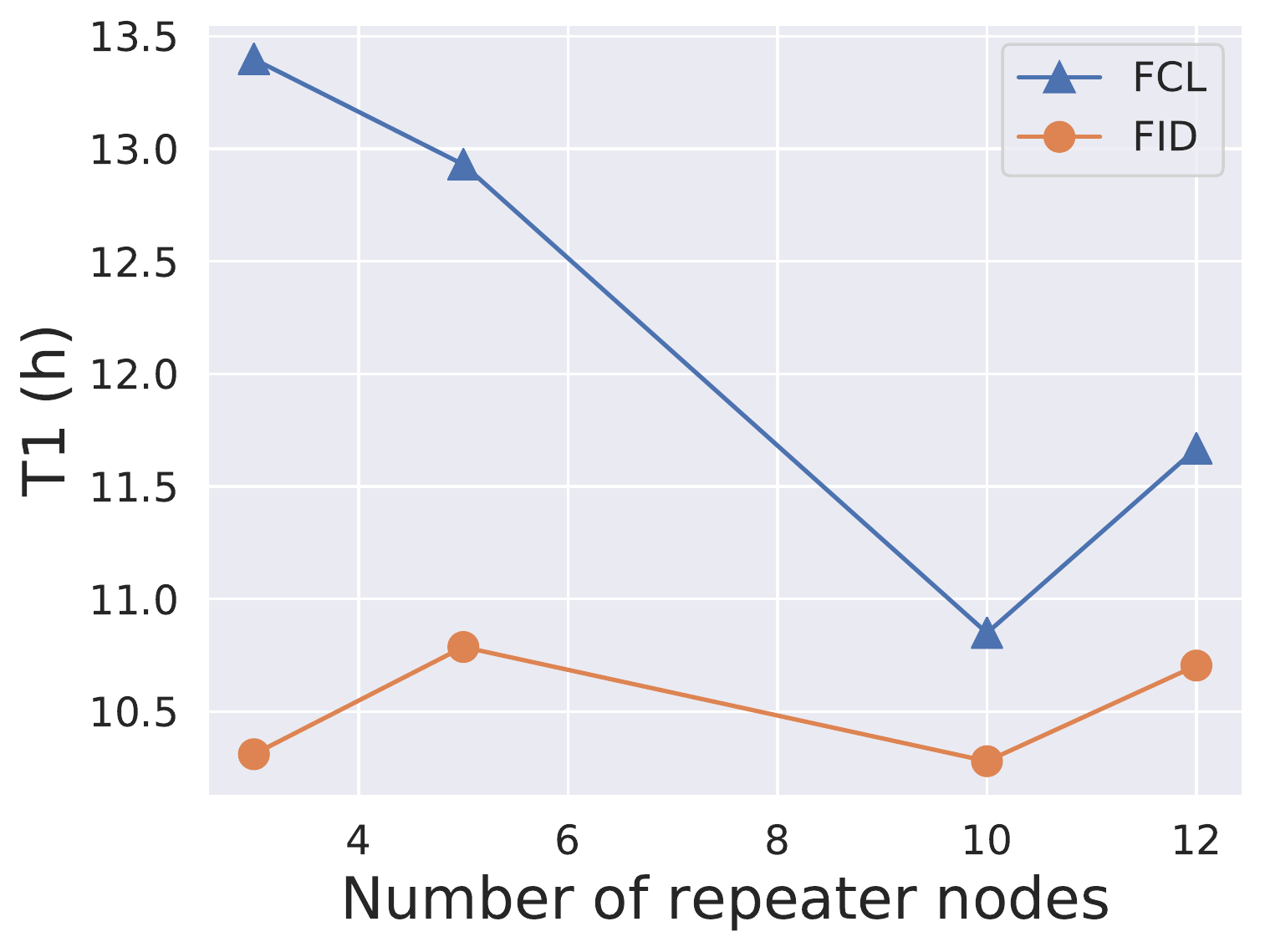} \\
    \small (a)
\end{tabular}
\begin{tabular}[b]{c}
    \includegraphics[width=0.32\textwidth]{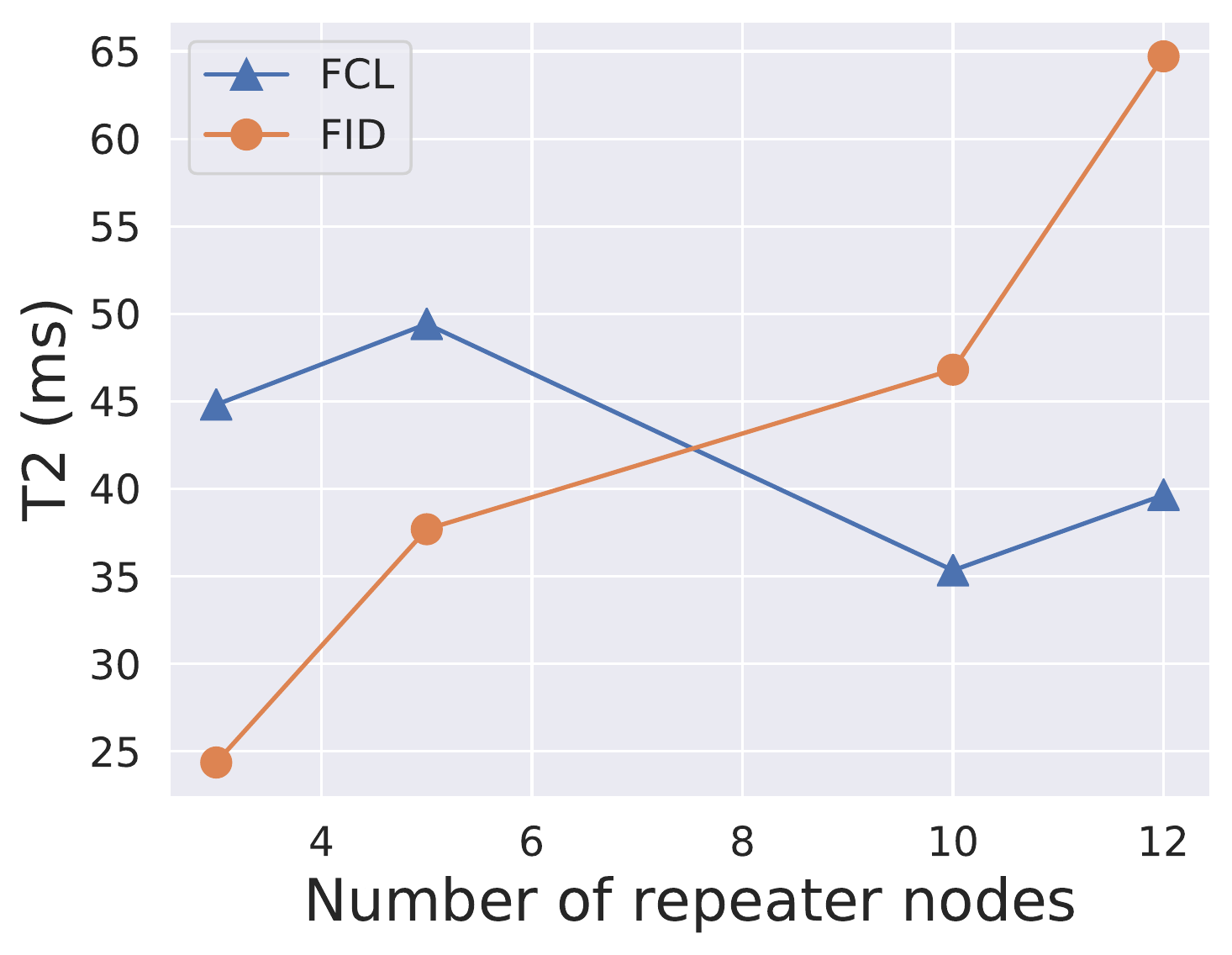} \\
    \small (b)
\end{tabular}
\begin{tabular}[b]{c}
    \includegraphics[width=0.32\textwidth]{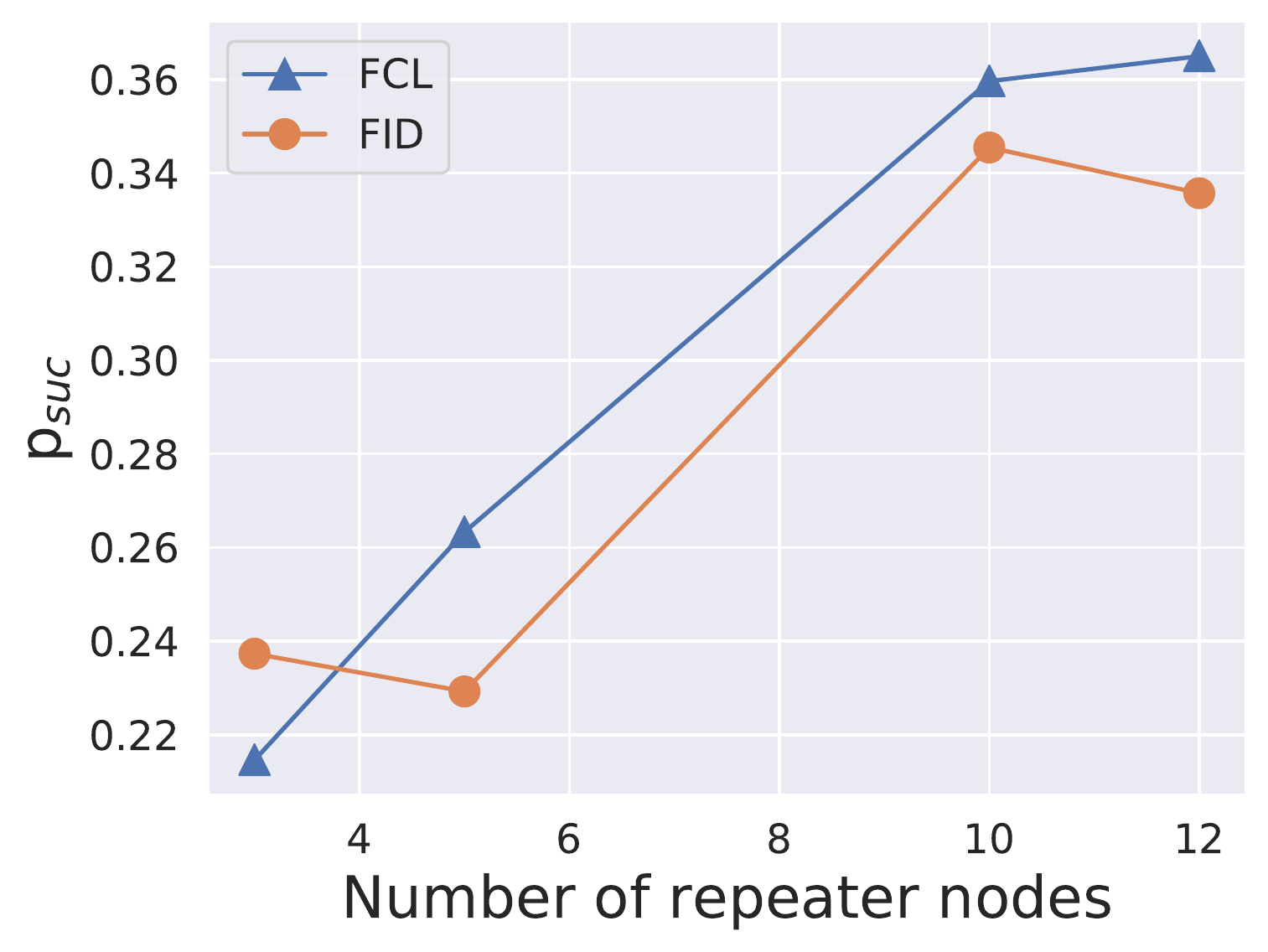} \\
    \small (c)
\end{tabular}
\begin{tabular}[b]{c}
    \includegraphics[width=0.32\textwidth]{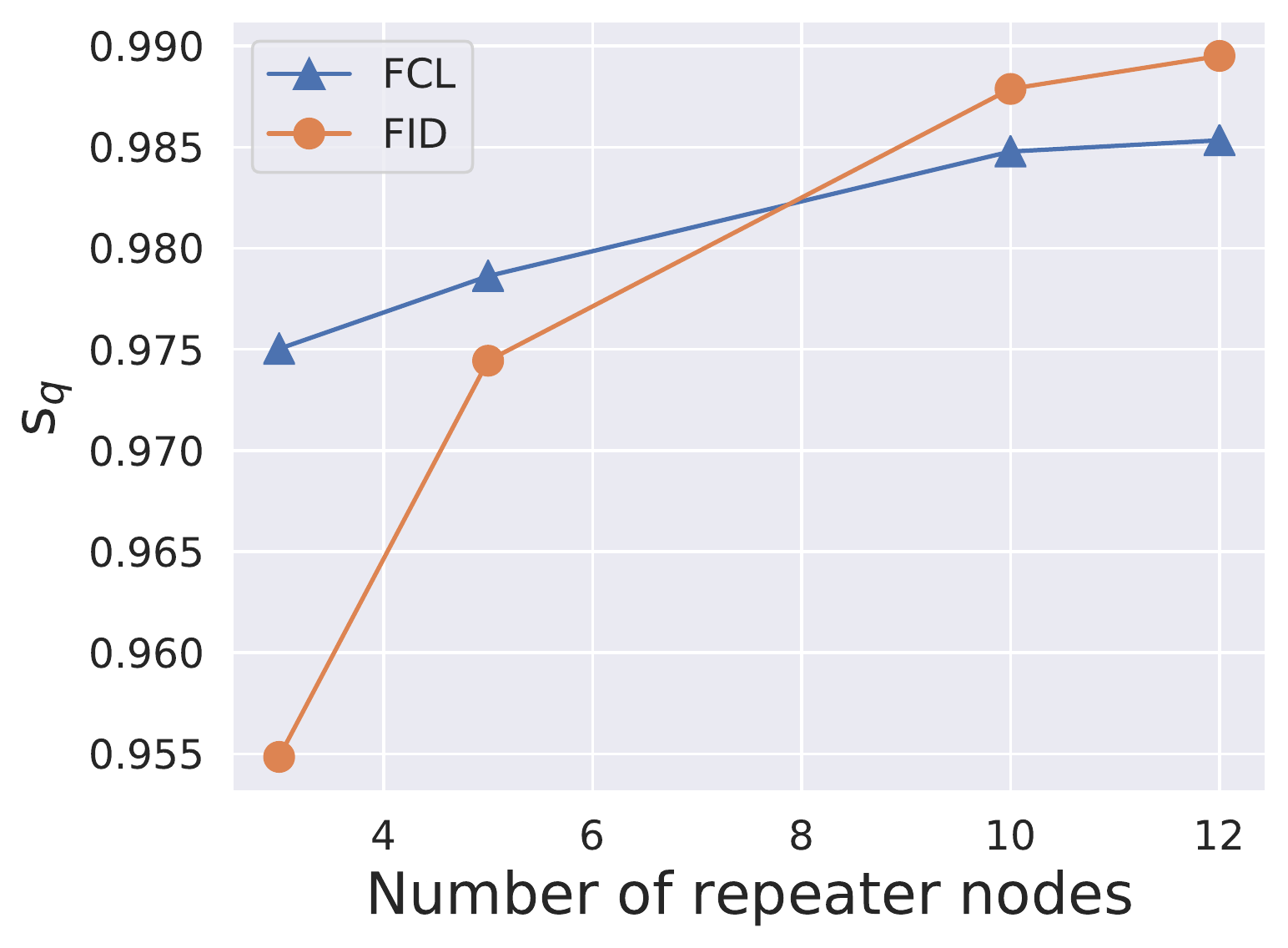} \\
    \small (d) 
\end{tabular}
\begin{tabular}[b]{c}
    \includegraphics[width=0.32\textwidth]{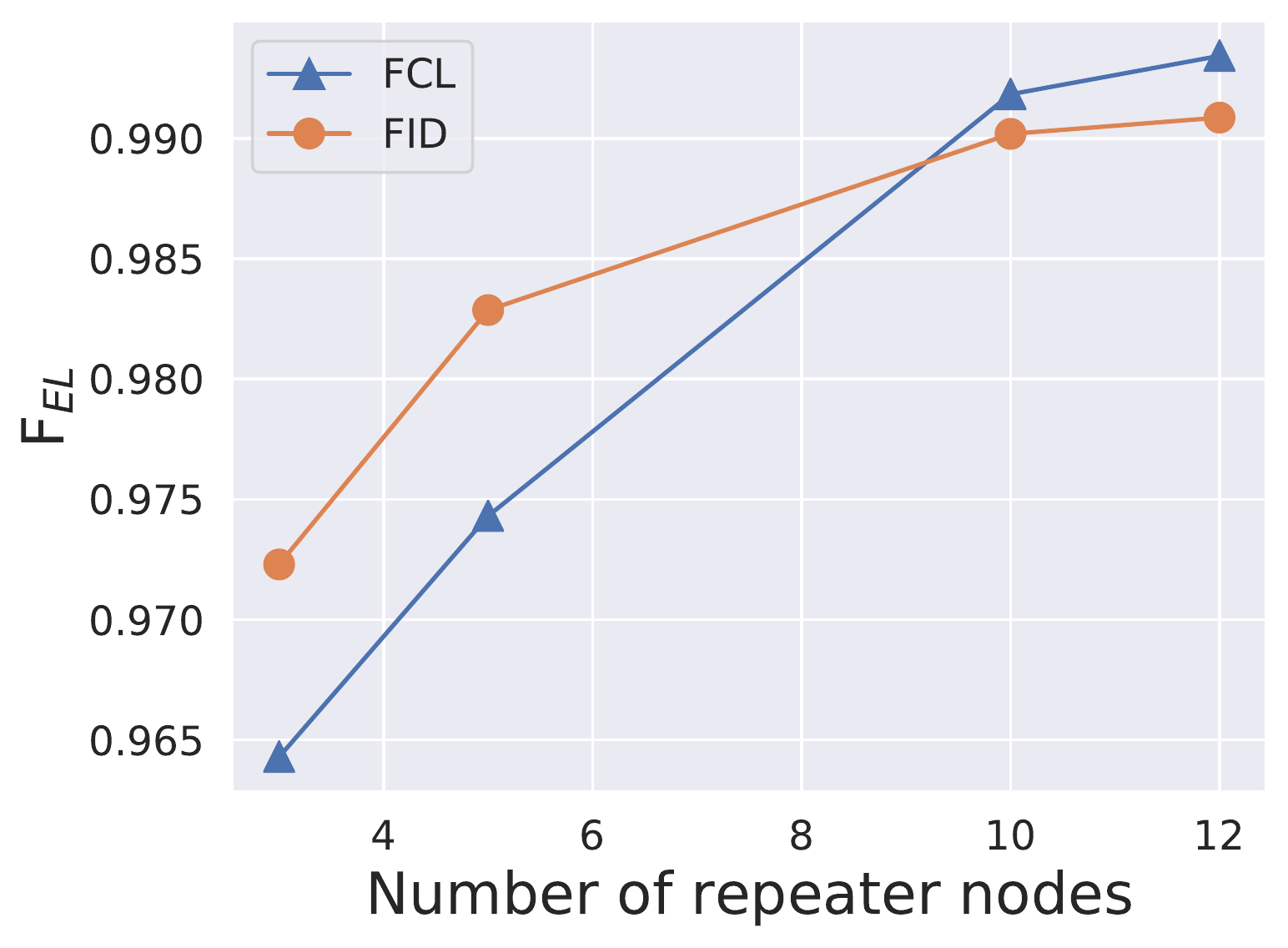} \\
    \small (e)
\end{tabular}

\caption{Parameters of the best solutions found for FCL and FID with different numbers of repeaters. Each data point corresponds to the best solution found after 200 generations, with 150 population individuals per generation and 100 simulation runs per individual. For a detailed discussion of these results, see the text in Section~\ref{sec:eq_spaced_nodes}.}
\label{fig:parameters_of_solutions}
\end{figure*}

To investigate why this happened, in Figure~\ref{fig:t2_psuc_best_solutions_both_setups} we plot the elementary link success probability $p_{suc}$ against the induced dephasing coherence time $T_2$. We can place the solutions found into two groups in terms of their $p_{suc}$ values, namely few-repeaters (to the left in the plot) and many-repeaters (to the right). In the few-repeaters group, solutions have comparatively lower $p_{suc}$ than in the many-repeaters group. In order to attain the same end-to-end entanglement generation rate over a chain with more nodes one needs a higher $p_{suc}$ as more elementary links must be generated. This is to some extent reflected in the grouping of solutions into few-repeaters and many-repeaters seen in the plot, but the fact that the optimal solutions attain different rates for different numbers of nodes somewhat confounds this analysis. In order to explain this, we note that $T_2$ and $p_{suc}$ are inextricably linked. $T_2$ reflects the intensity of the induced dephasing effect, (see Appendix~\ref{sec:validation}) with a higher value of $T_2$ corresponding to a weaker induced dephasing effect, and vice-versa. This type of noise is applied every time entanglement generation is attempted.  Therefore, its intensity heavily depends on $p_{suc}$: a lower success probability implies more entanglement generation attempts and thus more dephasing. One would naively think that the GA would always converge towards a solution with lower rate ($R$) up until the limit of $1$ Hz we defined, as that would allow for lower values of $p_{suc}$ and hence a lower value of the parameter cost. However, due to the connection between $p_{suc}$ and $T_2$, a lower value of the former necessitates a higher value of the latter. This then implies that solutions whose $R$ is closer to the established requirement of $1$ Hz, with their lower values of $p_{suc}$, might actually have higher costs than solutions with higher $R$, accounting for why the ideal solutions have such high rates.

To conclude our analysis of the solutions found with our optimization procedure, we present in Figure~\ref{fig:parameters_of_solutions} the values of each of the parameters of the solutions for each setup. Besides the relation between $p_{suc}$ and $T_2$, which we have already discussed, there are two interesting points to highlight. First, we note that the relative variations of $T_1$ for different setups are small when compared to the ones of $T_2$. Similarly to what we saw in the use case of Section~\ref{sec:real_network}, this indicates that $T_1$ is not a crucial parameter to improve for NV center-based repeaters. Second, both $F_{EL}$ and $s_q$ increase with the number of repeaters, approaching $1$. This was to be expected, as a higher number of repeaters implies more entanglement swaps and hence more decay in fidelity. Therefore, to reach the same end-to-end fidelity one needs better elementary links and swaps. 

We further note that for few repeaters, $F_{EL}$ is higher and $s_q$ is lower at FID than at FCL. The opposite is true for many repeaters. We believe this may be explained by the length of the elementary links in the FCL case. For few repeaters, the FCL elementary links are longer than the FID elementary links ($133-200$ km vs $100$ km), with the situation being reversed for many repeaters ($73-89$ km vs $100$ km). A longer elementary link translates into a worse baseline value of $F_{EL}$, as detailed in Appendix~\ref{sec:appendixbaseline}, and thus more expensive improvements. On the other hand, the baseline value of $s_q$ is the same irrespective of the elementary link length, and thus so is the cost of improving it. Therefore, for few repeaters the less costly solution at FCL has a lower elementary link fidelity and higher swap quality than the the less costly solution at FID. The opposite is true for many repeaters, explaining the observed behaviour.

\section{Conclusions} \label{sec:conclusions}
We have introduced a methodology for the optimization of entanglement generation and distribution in repeater chains using GAs. In contrast with previous work in this area \cite{wallnofer2020machine, jiang2007optimal, goodenough2020optimising, muralidharan2016optimal, santra2019quantum}, our methodology is systematic, modular and broadly applicable. We validated it by benchmarking our GAs on functions commonly used for this purpose and by applying it to a repeater chain generating Werner states. We can derive analytical results for such a chain and thus gauge how well our methodology performs. Having validated our methodology, we applied it to three use cases. First, we considered a repeater chain built using real-life fiber data, thus demonstrating that our methodology can go beyond simple network topologies. The other two use cases consisted of chains of equally spaced nodes for which we varied the number of repeaters. In one we kept the internode distance constant, and in the other we fixed the total chain length. By applying our methodology to these use cases we found what are the worst parameters achieving end-to-end fidelity and rate of at least $0.7$ and $1$ Hz, respectively, in different scenarios. Even though this was the question we focused on answering in this work, we must note that our methodology is more general and can be applied to a variety of problems, given that they can be restated as optimization problems and that an appropriate cost function is designed. 

On a similar note, we must again stress that even though we have here focused on a simplified five-parameter repeater model, in no way is our methodology restricted to such a model. In fact, one interesting application of our methodology would be to consider a more realistic hardware model, such as the one proposed in \cite{dahlberg2019link} for NV-center based repeaters. Such models are described by a very large number of parameters, on the order of $30$ in this case, which means that the initial search space is too large for a direct application of our methodology. To practically apply our methodology to such a large parameter space, one could opt for a two-stage optimization process. The first stage would be similar to what was shown in this work, i.e. applying the methodology to a simpler model that can be mapped to the more accurate one. This step would allow us to both reduce the search space by finding minimal requirements on parameters and to identify which of these parameters have a bigger impact on the target metrics. With this knowledge in hand, we could apply the methodology to a select subset of parameters in the more detailed model, performing the optimization procedure in a reduced, more feasible search space. The outcome of this two-step procedure would then be a realistic picture of what kind of hardware improvements are required to achieve long-range entanglement, constituting a useful guide for experimental groups working on repeater technology. This establishes the methodology we have proposed as an invaluable tool for the development of a blueprint for the quantum internet.

\section*{Acknowledgements} \label{sec:acknowledgements}
We thank Guus Avis, Kaushik Chakraborty, Axel Dahlberg, Hana Jirovská, Rob Knegjens and Julian Rabbie for insightful comments and discussions. We thank also Julian Rabbie and Matthew Skrzypczyk for critical reading of the manuscript. We further thank SURF for sharing data regarding their fiber network. This work was supported by the QIA-project that has received funding from the European Union’s Horizon 2020 research and innovation program under grant Agreement No. 820445, and also by an ERC Starting grant and NWO Zwaartekracht QSC.

\bibliography{biblio} 

\begin{thebibliography}{10}

\bibitem{bennett2020quantum}
C.~H. Bennett and G.~Brassard, ``Quantum cryptography: Public key distribution
  and coin tossing,'' {\em arXiv preprint arXiv:2003.06557}, 2020.

\bibitem{ekert1991quantum}
A.~K. Ekert, ``Quantum cryptography based on bell’s theorem,'' {\em Physical
  review letters}, vol.~67, no.~6, p.~661, 1991.

\bibitem{komar2014quantum}
P.~Komar, E.~M. Kessler, M.~Bishof, L.~Jiang, A.~S. S{\o}rensen, J.~Ye, and
  M.~D. Lukin, ``A quantum network of clocks,'' {\em Nature Physics}, vol.~10,
  no.~8, pp.~582--587, 2014.

\bibitem{buhrman2003distributed}
H.~Buhrman and H.~R{\"o}hrig, ``Distributed quantum computing,'' in {\em
  International Symposium on Mathematical Foundations of Computer Science},
  pp.~1--20, Springer, 2003.

\bibitem{wehner2018quantum}
S.~Wehner, D.~Elkouss, and R.~Hanson, ``Quantum internet: A vision for the road
  ahead,'' {\em Science}, vol.~362, no.~6412, 2018.

\bibitem{hensen2015loophole}
B.~Hensen, H.~Bernien, A.~E. Dr{\'e}au, A.~Reiserer, N.~Kalb, M.~S. Blok,
  J.~Ruitenberg, R.~F. Vermeulen, R.~N. Schouten, C.~Abell{\'a}n, {\em et~al.},
  ``Loophole-free bell inequality violation using electron spins separated by
  1.3 kilometres,'' {\em Nature}, vol.~526, no.~7575, pp.~682--686, 2015.

\bibitem{briegel1998quantum}
H.-J. Briegel, W.~D{\"u}r, J.~I. Cirac, and P.~Zoller, ``Quantum repeaters: the
  role of imperfect local operations in quantum communication,'' {\em Physical
  Review Letters}, vol.~81, no.~26, p.~5932, 1998.

\bibitem{rozpkedek2019near}
F.~Rozp{\k{e}}dek, R.~Yehia, K.~Goodenough, M.~Ruf, P.~C. Humphreys, R.~Hanson,
  S.~Wehner, and D.~Elkouss, ``Near-term quantum-repeater experiments with
  nitrogen-vacancy centers: Overcoming the limitations of direct
  transmission,'' {\em Physical Review A}, vol.~99, no.~5, p.~052330, 2019.

\bibitem{humphreys2018deterministic}
P.~C. Humphreys, N.~Kalb, J.~P. Morits, R.~N. Schouten, R.~F. Vermeulen, D.~J.
  Twitchen, M.~Markham, and R.~Hanson, ``Deterministic delivery of remote
  entanglement on a quantum network,'' {\em Nature}, vol.~558, no.~7709,
  pp.~268--273, 2018.

\bibitem{yu2020entanglement}
Y.~Yu, F.~Ma, X.-Y. Luo, B.~Jing, P.-F. Sun, R.-Z. Fang, C.-W. Yang, H.~Liu,
  M.-Y. Zheng, X.-P. Xie, {\em et~al.}, ``Entanglement of two quantum memories
  via fibres over dozens of kilometres,'' {\em Nature}, vol.~578, no.~7794,
  pp.~240--245, 2020.

\bibitem{zwerger2017quantum}
M.~Zwerger, B.~Lanyon, T.~Northup, C.~Muschik, W.~D{\"u}r, and N.~Sangouard,
  ``Quantum repeaters based on trapped ions with decoherence-free subspace
  encoding,'' {\em Quantum Science and Technology}, vol.~2, no.~4, p.~044001,
  2017.

\bibitem{munro2015inside}
W.~J. Munro, K.~Azuma, K.~Tamaki, and K.~Nemoto, ``Inside quantum repeaters,''
  {\em IEEE Journal of Selected Topics in Quantum Electronics}, vol.~21, no.~3,
  pp.~78--90, 2015.

\bibitem{sangouard2011quantum}
N.~Sangouard, C.~Simon, H.~De~Riedmatten, and N.~Gisin, ``Quantum repeaters
  based on atomic ensembles and linear optics,'' {\em Reviews of Modern
  Physics}, vol.~83, no.~1, p.~33, 2011.

\bibitem{netsquid}
``Netsquid.'' \url{https://netsquid.org}.

\bibitem{coopmans2020netsquid}
T.~Coopmans, R.~Knegjens, A.~Dahlberg, D.~Maier, L.~Nijsten, J.~Oliveira,
  M.~Papendrecht, J.~Rabbie, F.~Rozpędek, M.~Skrzypczyk, L.~Wubben,
  W.~de~Jong, D.~Podareanu, A.~T. Knoop, D.~Elkouss, and S.~Wehner, ``Netsquid,
  a discrete-event simulation platform for quantum networks,'' 2020.

\bibitem{van2008system}
R.~Van~Meter, T.~D. Ladd, W.~J. Munro, and K.~Nemoto, ``System design for a
  long-line quantum repeater,'' {\em IEEE/ACM Transactions on Networking},
  vol.~17, no.~3, pp.~1002--1013, 2008.

\bibitem{wu2020sequence}
X.~Wu, A.~Kolar, J.~Chung, D.~Jin, T.~Zhong, R.~Kettimuthu, and M.~Suchara,
  ``Sequence: A customizable discrete-event simulator of quantum networks,''
  {\em arXiv preprint arXiv:2009.12000}, 2020.

\bibitem{wallnofer2020machine}
J.~Walln{\"o}fer, A.~A. Melnikov, W.~D{\"u}r, and H.~J. Briegel, ``Machine
  learning for long-distance quantum communication,'' {\em PRX Quantum},
  vol.~1, no.~1, p.~010301, 2020.

\bibitem{muralidharan2016optimal}
S.~Muralidharan, L.~Li, J.~Kim, N.~L{\"u}tkenhaus, M.~D. Lukin, and L.~Jiang,
  ``Optimal architectures for long distance quantum communication,'' {\em
  Scientific reports}, vol.~6, p.~20463, 2016.

\bibitem{jiang2007optimal}
L.~Jiang, J.~M. Taylor, N.~Khaneja, and M.~D. Lukin, ``Optimal approach to
  quantum communication using dynamic programming,'' {\em Proceedings of the
  National Academy of Sciences}, vol.~104, no.~44, pp.~17291--17296, 2007.

\bibitem{santra2019quantum}
S.~Santra, L.~Jiang, and V.~S. Malinovsky, ``Quantum repeater architecture with
  hierarchically optimized memory buffer times,'' {\em Quantum Science and
  Technology}, vol.~4, no.~2, p.~025010, 2019.

\bibitem{goodenough2020optimising}
K.~Goodenough, D.~Elkouss, and S.~Wehner, ``Optimising repeater schemes for the
  quantum internet,'' {\em arXiv preprint arXiv:2006.12221}, 2020.

\bibitem{krastanov2019optimized}
S.~Krastanov, V.~V. Albert, and L.~Jiang, ``Optimized entanglement
  purification,'' {\em Quantum}, vol.~3, p.~123, 2019.

\bibitem{smartstopos}
A.~Torres-Knoop, T.~Coopmans, D.~Maier, and F.~Silva, ``smart-stopos.''
  \url{https://gitlab.com/aritoka/smart-stopos}, 2020.

\bibitem{werner1989quantum}
R.~F. Werner, ``Quantum states with einstein-podolsky-rosen correlations
  admitting a hidden-variable model,'' {\em Physical Review A}, vol.~40, no.~8,
  p.~4277, 1989.

\bibitem{gunantara2018review}
N.~Gunantara, ``A review of multi-objective optimization: Methods and its
  applications,'' {\em Cogent Engineering}, vol.~5, no.~1, p.~1502242, 2018.

\bibitem{schaffer1986some}
J.~D. Schaffer, ``Some experiments in machine learning using vector evaluated
  genetic algorithms (artificial intelligence, optimization, adaptation,
  pattern recognition),'' 1986.

\bibitem{collins2007multiplexed}
O.~Collins, S.~Jenkins, A.~Kuzmich, and T.~Kennedy, ``Multiplexed
  memory-insensitive quantum repeaters,'' {\em Physical review letters},
  vol.~98, no.~6, p.~060502, 2007.

\bibitem{vikhar2016evolutionary}
P.~A. Vikhar, ``Evolutionary algorithms: A critical review and its future
  prospects,'' in {\em 2016 International conference on global trends in signal
  processing, information computing and communication (ICGTSPICC)},
  pp.~261--265, IEEE, 2016.

\bibitem{holland1992adaptation}
J.~H. Holland {\em et~al.}, {\em Adaptation in natural and artificial systems:
  an introductory analysis with applications to biology, control, and
  artificial intelligence}.
\newblock MIT press, 1992.

\bibitem{beyer2002evolution}
H.-G. Beyer and H.-P. Schwefel, ``Evolution strategies--a comprehensive
  introduction,'' {\em Natural computing}, vol.~1, no.~1, pp.~3--52, 2002.

\bibitem{storn1997differential}
R.~Storn and K.~Price, ``Differential evolution--a simple and efficient
  heuristic for global optimization over continuous spaces,'' {\em Journal of
  global optimization}, vol.~11, no.~4, pp.~341--359, 1997.

\bibitem{kennedy1995particle}
J.~Kennedy and R.~Eberhart, ``Particle swarm optimization,'' in {\em
  Proceedings of ICNN'95-International Conference on Neural Networks}, vol.~4,
  pp.~1942--1948, IEEE, 1995.

\bibitem{shi1998modified}
Y.~Shi and R.~Eberhart, ``A modified particle swarm optimizer,'' in {\em 1998
  IEEE international conference on evolutionary computation proceedings. IEEE
  world congress on computational intelligence (Cat. No. 98TH8360)},
  pp.~69--73, IEEE, 1998.

\bibitem{goldberg2006genetic}
D.~E. Goldberg, {\em Genetic algorithms}.
\newblock Pearson Education India, 2006.

\bibitem{stopos}
``stopos.'' \url{https://gitlab.com/surfsara/stopos}, 2019.

\bibitem{digalakis2001benchmarking}
J.~G. Digalakis and K.~G. Margaritis, ``On benchmarking functions for genetic
  algorithms,'' {\em International journal of computer mathematics}, vol.~77,
  no.~4, pp.~481--506, 2001.

\bibitem{10.5555/534133}
D.~E. Goldberg, {\em Genetic Algorithms in Search, Optimization and Machine
  Learning}.
\newblock USA: Addison-Wesley Longman Publishing Co., Inc., 1st~ed., 1989.

\bibitem{bradley2019ten}
C.~Bradley, J.~Randall, M.~Abobeih, R.~Berrevoets, M.~Degen, M.~Bakker,
  M.~Markham, D.~Twitchen, and T.~Taminiau, ``A ten-qubit solid-state spin
  register with quantum memory up to one minute,'' {\em Physical Review X},
  vol.~9, no.~3, p.~031045, 2019.

\bibitem{rabbie2020designing}
J.~Rabbie, K.~Chakraborty, G.~Avis, and S.~Wehner, ``Designing quantum networks
  using preexisting infrastructure,'' {\em arXiv preprint arXiv:2005.14715},
  2020.

\bibitem{dahlberg2019link}
A.~Dahlberg, M.~Skrzypczyk, T.~Coopmans, L.~Wubben, F.~Rozp{\k{e}}dek,
  M.~Pompili, A.~Stolk, P.~Pawe{\l}czak, R.~Knegjens, J.~de~Oliveira~Filho,
  {\em et~al.}, ``A link layer protocol for quantum networks,'' in {\em
  Proceedings of the ACM Special Interest Group on Data Communication},
  pp.~159--173, 2019.

\bibitem{srinivas1994adaptive}
M.~Srinivas and L.~M. Patnaik, ``Adaptive probabilities of crossover and
  mutation in genetic algorithms,'' {\em IEEE Transactions on Systems, Man, and
  Cybernetics}, vol.~24, no.~4, pp.~656--667, 1994.

\bibitem{Luke2013Metaheuristics}
S.~Luke, {\em Essentials of Metaheuristics}.
\newblock Lulu, second~ed., 2013.
\newblock Available for free at
  http://cs.gmu.edu/$\sim$sean/book/metaheuristics/.

\bibitem{kalb2017entanglement}
N.~Kalb, A.~A. Reiserer, P.~C. Humphreys, J.~J. Bakermans, S.~J. Kamerling,
  N.~H. Nickerson, S.~C. Benjamin, D.~J. Twitchen, M.~Markham, and R.~Hanson,
  ``Entanglement distillation between solid-state quantum network nodes,'' {\em
  Science}, vol.~356, no.~6341, pp.~928--932, 2017.

\bibitem{kalb2018dephasing}
N.~Kalb, P.~C. Humphreys, J.~Slim, and R.~Hanson, ``Dephasing mechanisms of
  diamond-based nuclear-spin memories for quantum networks,'' {\em Physical
  Review A}, vol.~97, no.~6, p.~062330, 2018.

\bibitem{dur1999quantum}
W.~D{\"u}r, H.-J. Briegel, J.~I. Cirac, and P.~Zoller, ``Quantum repeaters
  based on entanglement purification,'' {\em Physical Review A}, vol.~59,
  no.~1, p.~169, 1999.

\bibitem{wales1997global}
D.~J. Wales and J.~P. Doye, ``Global optimization by basin-hopping and the
  lowest energy structures of lennard-jones clusters containing up to 110
  atoms,'' {\em The Journal of Physical Chemistry A}, vol.~101, no.~28,
  pp.~5111--5116, 1997.

\bibitem{privatecomm}
S.~Hermans. private communication, 2020.

\end{thebibliography}
\bibliographystyle{ieeetr}

\appendix*
\appendix
\section{smart-stopos}\label{sec:appendixsmartstopos}
\begin{figure*}[!ht]
\centering
\includegraphics[width=0.75\textwidth]{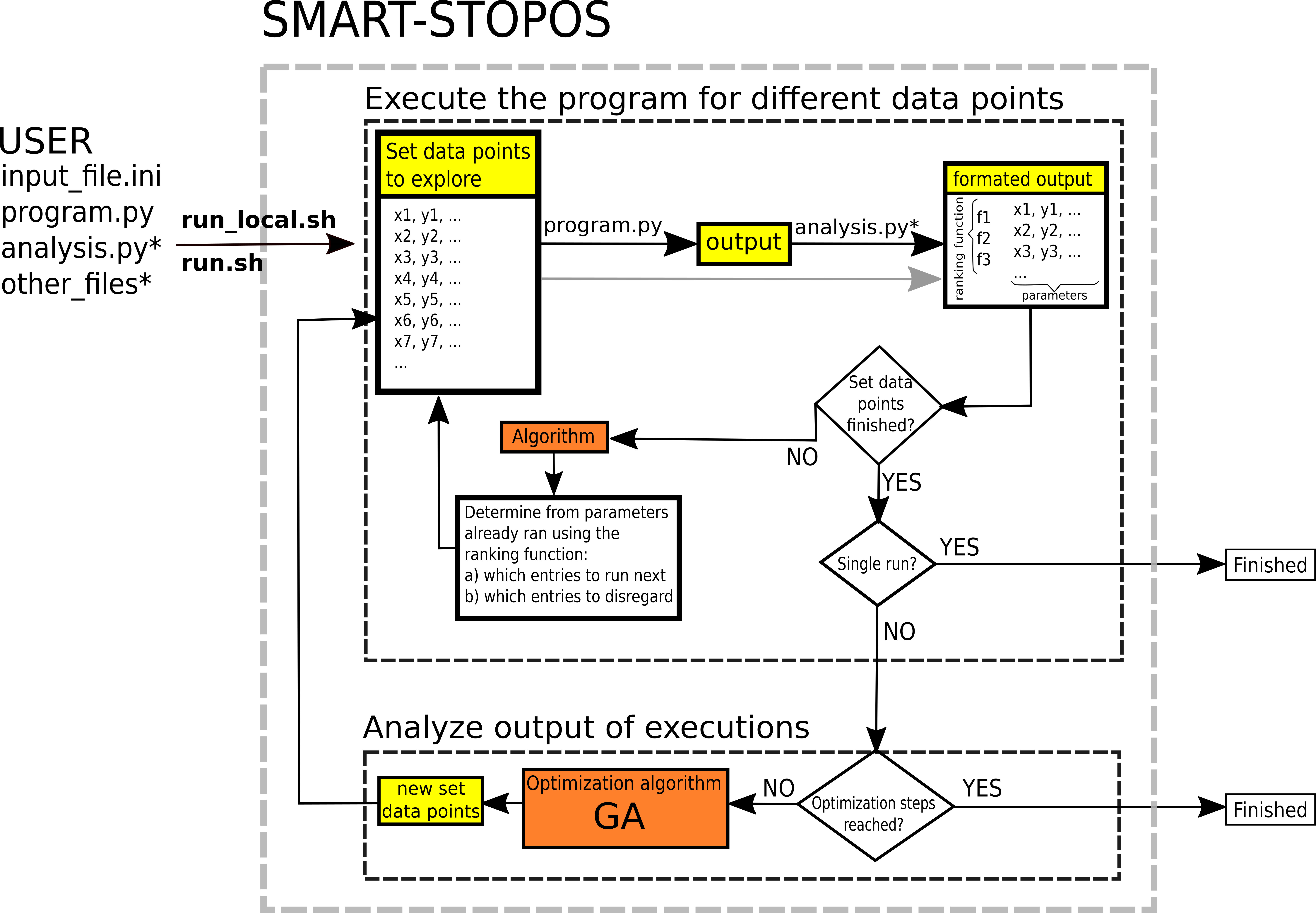}
\caption{Diagram of the \textit{smart-stopos} workflow for parameter optimization. input\_file.ini is used to define optimization parameters such as algorithm to be used, parameters to optimize and allowed range of values. The execution of a simulation (program.py) and the analysis of resulting data (analysis.py) can be done locally (run\_local.sh) or using HPC facilities (run.sh). The stopos~\cite{stopos} job manager tool is required for running on HPC facilities. Files marked with * are optional.}
\label{fig:smartstopos}
\end{figure*}

 In Figure~\ref{fig:smartstopos}, we present a detailed overview of the \textit{smart-stopos} workflow. The user must provide a script, entitled program.py in Figure~\ref{fig:smartstopos}, that runs the simulation and an input\_file.ini that contains information about the optimization procedure, such as the number of iterations and parameter specifications. Given these inputs, \textit{smart-stopos} generates sets of parameters for which the simulation will be run according to the specifications given in input\_file.ini. The outputs of the simulation are then used to generate a new set of parameters for the next iteration. This generation is done in an algorithm-dependent way. We used GAs in this work but in principle any other algorithm could be plugged in, provided that it can be run with only simulation inputs and outputs.

\section{Genetic Algorithms}\label{sec:appendiGA}
In this Appendix we give a detailed view of the GA implementation we used for the simulations described in this work.

We started by transforming all parameters to be in the $[0,1]$ range. This is trivial for the elementary link fidelity, success probability and swap quality. For $T_1$ and $T_2$, which usually live in the $[0,\infty]$ range, we performed the following transformation:
\begin{equation}
    T' = \begin{cases} \frac{1}{T + 1} & \quad \text{if T > 0}\\
    0                                  & \quad \text{o.w.},
    \end{cases}
\end{equation}
which results in $T' \in [0,1]$, as required. A chromosome, i.e. a set of parameters constituting a candidate solution, is thus a set of 5 real numbers in the $[0,1]$ interval.

We used populations of 150 individuals, as the literature suggests that numbers of this order of magnitude are enough to get adequate parameter space exploration while still being computationally feasible \cite{goldberg2006genetic}. 

After the cost function is computed for all members of the population, we select 10 of them, 20\% of the total population, according to the Roulette Wheel method \cite{goldberg2006genetic}. Again, the literature indicates that the percentage of selected individuals should be of this order of magnitude and we empirically verified that this value produced the best results for our particular use case. One of the major challenges in GA-based optimization is to balance exploration of the search space with exploitation of known minima. If the algorithm performs selection in a purely random manner, it is no different than random search. On the other hand, if it simply selects the best individuals in a given generation, the population will tend to get stuck in local minima and be vulnerable to premature convergence. The Roulette Wheel selection method is a well-known approach to this problem, balancing exploration and exploitation by assigning selection probabilities to individuals biased, but not completely determined, by their fitness value. Applying this method to a maximization problem, the probability $p_i$ of individual $i$ being selected is given by:
\begin{equation}
    p_i = \frac{f_i}{\sum_j f_j},
\end{equation}
with $f_j$ being the value of the fitness function for individual $j$. The probability of selection is then proportional to how a big of a share of the total fitness the individual's fitness represents, i.e. how good it is in comparison to its peers. Our problem is, however, one of minimization, not maximization. Therefore, we adapted this method by simply inverting the values of the fitness function.

Crossover is subsequently applied on the 10 selected members of the population, known as parents. This is done by randomly choosing two of the parents, sampling a crossover point, and mixing the two accordingly. To give a concrete example, if the chromosomes of the two parents are given by $[a_1, a_2, a_3, a_4, a_5]$ and $[b_1, b_2, b_3, b_4, b_5]$ and the crossover point was $2$, the resulting child would have chromosome $[a_1, a_2, b_3, b_4, b_5]$. The number of children generated in this way is given by the crossover parameter, a hyperparameter of the algorithm defining how often crossover happens, times the desired population size.  

The parents plus the children resulting from the crossover process are then mutated. In this process, all chromosomes of a given member of the population are randomly changed by some value that keeps them inside their range. For the mutation probability of a given parent, we implemented the adaptive scheme introduced in \cite{srinivas1994adaptive}, which was shown the reduce the likelihood of corrupting a high-quality solution and enhance the exploratory properties of the algorithm. In this scheme, the probability of parent $k$ being mutated is given by:
\begin{equation}
    p_m = \begin{cases}
    0.5 & \quad \text{if c$_k$ > $\Bar{\text{c}}$} \\
    0.5\frac{\text{c}_k - \text{c}_{\text{min}}}{\Bar{\text{c}} - \text{c}_{\text{min}}} & \quad \text{o.w.},
    \end{cases}
\end{equation}
where $\text{c}_k$ is the value of the cost function for parent $k$, $\Bar{\text{c}}$ is the value of the cost function averaged over the previous generation's population and $\text{c}_{\text{min}}$ is its minimum value. For the children generated in the crossover process, for which there is no cost value yet, the mutation probability is a hyperparameter of the algorithm. Previous work suggests that a high cross over parameter and low mutation probability produce good results \cite{goldberg2006genetic}, so we used a crossover parameter of $0.7$ and a mutation probability of $0.02$ to obtain the results showed in this work. 

Since generation of new individuals is to some extent probabilistic, the size of a generation can vary. To keep our population size fixed, we either randomly remove elements or add some of the best members of the previous generation. We also implement a form of elitism, meaning that the best element of the previous generation is always preserved in the following generation, in order to prevent the algorithm wasting time searching for solutions it has already found \cite{Luke2013Metaheuristics}.

We have empirically determined that $200$ generations are usually enough to achieve satisfying solutions while still being computationally feasible on a cluster.

\section{Abstract Model Validation}
In this Appendix we show how we validated the abstract model against a physically-accurate NV model.

\label{sec:am_validation}
\subsection{Matching to NV Model}
In order to ensure that the simulations of the abstract model can contribute to our understanding of actual physical implementations of quantum repeaters, we must verify that this abstract model captures the relevant physics to a reasonable extent. To do so, we will compare the results of simulations of a repeater chain in the abstract model with those of a repeater chain running a physically accurate model. For this purpose, any model of a physical system being studied as a possible platform for quantum repeaters would do. We will thus focus on one such system, namely NV centers, modelled as described in \cite{kalb2017entanglement}. This is a very detailed model that accurately captures the physics of NV centers, including for instance modelling the photon emission, capture and detection processes as well as differentiating between communication and memory qubits, with all the restrictions that entails. In contrast, the simplified model we consider abstracts away all of the subtleties of photon emission and detection into an overarching success probability and treats all qubits as equal. Another key difference is that in the NV model the parameters are not mutually independent e.g. there is a relation of inverse proportionality between the fidelity of the generated entangled states and the rate at which they are generated due to the fact that both of these parameters depend on the bright state population. On the other hand, in the abstract model we make the simplifying assumption that all parameters are independent from one another. We must however emphasize that this does not reflect a limitation of our method. Taking the constraints arising from interparameter dependence into account would be possible, but we chose not to consider any such constraints in this preliminary study. 

More concretely, we will perform the validation of the abstract model by taking a set of parameters describing an NV center in the model, converting it to the five parameter set that defines our model, running both simulations, and checking how the end-to-end fidelity and entanglement generation rate compare. 

We start by proposing a mapping from the NV model in \cite{kalb2017entanglement} to the five-parameter abstract model we introduced in \ref{sec:abstractmodel}. We assume that elementary link states generated in the abstract repeater chain are of the form:

\begin{equation}
    \ket{\phi}\bra{\phi} = F_{EL} \ket{\psi}\bra{\psi} + (1 - F_{EL}) \ket{\uparrow\uparrow}\bra{\uparrow\uparrow},
\label{eq:abstract_states}
\end{equation}
where $\ket{\psi}\bra{\psi}$ is the ideal Bell state, $F_{EL}$ is the elementary link fidelity and $\ket{\uparrow\uparrow}\bra{\uparrow\uparrow}$ is given by:

\begin{equation*}
    \ket{\uparrow\uparrow}\bra{\uparrow\uparrow} = \begin{bmatrix}
    1 & 0 & 0 & 0 \\
    0 & 0 & 0 & 0 \\
    0 & 0 & 0 & 0 \\
    0 & 0 & 0 & 0 \\
    \end{bmatrix}.
\end{equation*}
The overlap between $\ket{\psi}$ and $\ket{\uparrow\uparrow}$ is 0, so $F_{EL}$ is in fact the elementary link fidelity, the sole parameter defining elementary link states. To map states from one model to another we compute the fidelity of the NV state described in the appendix of \cite{kalb2017entanglement} and use the result to define the abstract model state as in Equation \eqref{eq:abstract_states}. The probability of successfully generating these elementary links is obtained in an identical manner.

We take into account any errors that might occur in an entanglement swap, which include gate errors, measurement errors and initialization errors by modelling them all as depolarizing channels, with parameters $\{p_i\}$ and multiplying them to obtain a single parameter, $s_q$, as shown in Equation \eqref{eq:swap_quality}.

\begin{equation}
    s_q = \prod_i (1 - p_i)
\label{eq:swap_quality}
\end{equation}
$1 - s_q$ is then used to parameterize a depolarizing channel that is applied after an ideal Bell state measurement. The action of this channel $\Phi$ on a given state $\rho$ as a function of $s_q$ is given by 
\begin{equation}
\footnotesize
    \Phi(\rho, s_q) = \left(\frac{1 + 3 s_q}{4}\right) \rho + \frac{1 - s_q}{4} (X\rho X + Y\rho Y + Z \rho Z).   
\label{eq:depol_channel}
\end{equation}
This implies that $s_q$ is a measure of the quality of an entanglement swap, and it is thus named swap quality.

The two remaining parameters in the abstract model are $T_1$ and $T_2$. An NV center's qubits can be either electrons, used as communication qubits, or carbons, used as memory qubits, each of them having different coherence times. This subtlety is lost when going to the abstract model, in which all qubits are created equal. We expect that decoherence will be more relevant in the memory qubits than in the communication qubits, so we ignore it for the latter. Besides this, one of the major sources of noise in NV centers is induced dephasing, the dephasing applied to the memory qubits whenever the communication qubit attempts to generate entanglement \cite{kalb2018dephasing}. This noise source can also be accurately modelled by a $T_1$, $T_2$ noise model. In such a model, one applies dephasing noise with probability given by
\begin{equation}
    p = \frac{1 - e^{-t(1/T_2 - 1/2T_1)}}{2},
    \label{eq:prob_dephasing_T2}
\end{equation}
with $t$ being the relevant time period. This is formalized by means of a dephasing channel $\Phi_d$ whose action on a given state $\rho$ is given by
\begin{equation}
    \Phi_d (\rho, p) = (1-p) \rho + p Z\rho Z.
\end{equation}
On the other hand, the noise introduced in a NV center's carbon atoms over $n$ entanglement generation attempts can be modelled by a dephasing noise process of probability
\begin{equation}
    p_n = \frac{1 + \left(2 (1 - p_1) - 1\right)^n}{2},
\label{eq:prob_dephasing_n_attempts}
\end{equation}
with $p_1$ being the probability of a single attempt inducing dephasing noise, which can be experimentally determined \cite{kalb2018dephasing}. If we assume that a node is always trying to generate entanglement through its electron, we can write $n$ as a function of time:
\begin{equation}
    n = \frac{t}{T_{cycle}},
\end{equation}
with $T_{cycle}$ being the time it takes the NV to go through one entanglement generation attempt. 

Matching the probability in Equation~\eqref{eq:prob_dephasing_T2} to the one in Equation~\eqref{eq:prob_dephasing_n_attempts} and solving for $T_2$, we find:
\begin{equation}
    T_2 = \frac{1}{1/2T_1 - \log(1 - 2p_1)/T_{cycle}}.
\label{eq:T_2_induced_carbon_decay}    
\end{equation}
This allows us to account for the effect of induced dephasing in our simulations by modelling it as a $T_2$ noise process. We note that, in order to more closely capture induced dephasing, this noise should only be simulated when nodes are attempting entanglement generation.

In summary, we have two important sources of noise that can be modelled by $T_1$, $T_2$ processes: induced dephasing and memory decoherence. Since we want to restrict our model to $5$ parameters, we must restrict ourselves to account for one of the two. In order to make an informed decision regarding which noise source to model, we run repeater chain simulations using the abstract model and the NV model introduced in \cite{kalb2017entanglement}. For simplicity, we ignore distillation and consider a SWAP-ASAP protocol where the nodes can only attempt entanglement generation, wait or perform an entanglement swap. In order to obtain a better agreement between the entanglement generation rates of both models, we impose that nodes in the abstract model simulation can only generate entanglement with one neighbour at a time, as is the case for NV centers.

\subsection{Comparison of NV and Abstract Models}
We will look into how the internode distance affects the metrics we are interested in, namely end-to-end fidelity and entanglement generation rate, in the two models. To do so, we will focus on chains of equally spaced nodes, for which varying the internode distance is equivalent to varying the total length of the repeater chain. 

\begin{figure}[!ht]
\centering
\includegraphics[width=\columnwidth]{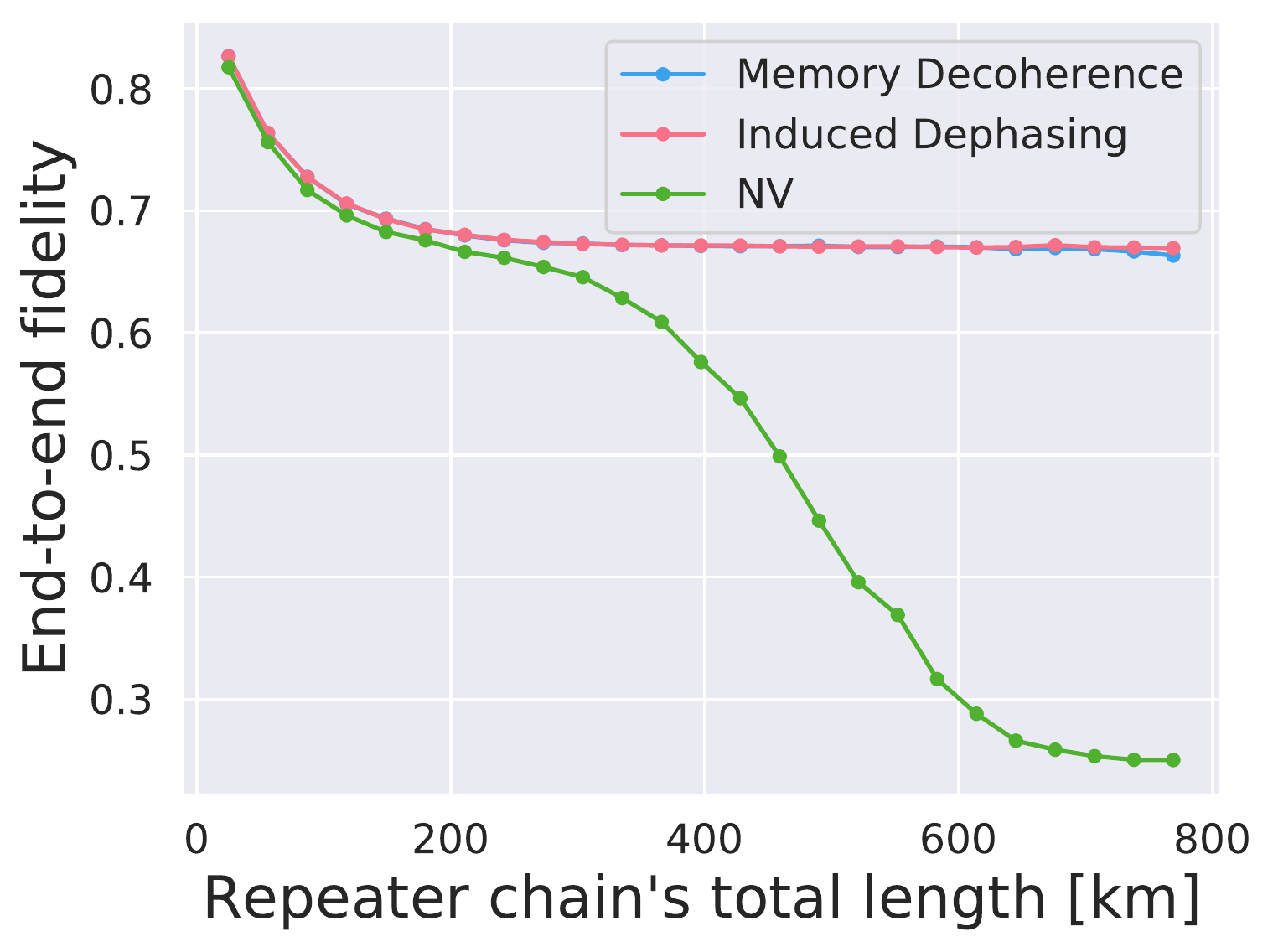}
\caption{Variation of the end-to-end fidelity of states generated by a chain of five equally spaced nodes as the chain's length is varied in the NV model (green, triangles) and in both abstract model mappings: memory decoherence (blue, circles) and induced dephasing (red, inverted triangles). The curves overlap for short chains, but as the internode distance grows the fidelity of the NV chain falls faster. The results of the two mappings are virtually identical. The error bars are smaller than the markers.}
\label{fig:validate_abstract_5nodes_fidelity_distance}
\end{figure}

In Figure~\ref{fig:validate_abstract_5nodes_fidelity_distance}, we plot the end-to-end fidelity of the states generated by a chain of five equally spaced nodes as the chain's length is varied in the NV model and in both abstract model mappings.
The three fidelity curves are very similar for shorter chains, roughly overlapping in chains of up to 200 km. At this point the curve for the NV model starts to diverge, dropping abruptly.

Overall, the difference between the two mappings is small. They both show very good agreement at short chain lengths, and they both perform poorly as the distances grows. This indicates that for longer distances or, alternatively, for lower fidelities, ignoring either of the noise sources results in poor agreement with the NV model. In this work we will focus on scenarios where the obtained fidelities are high, above 0.7, so that the agreement is good. We will consider the induced dephasing mapping.

\begin{figure}[!ht]
\centering
\includegraphics[width=\columnwidth]{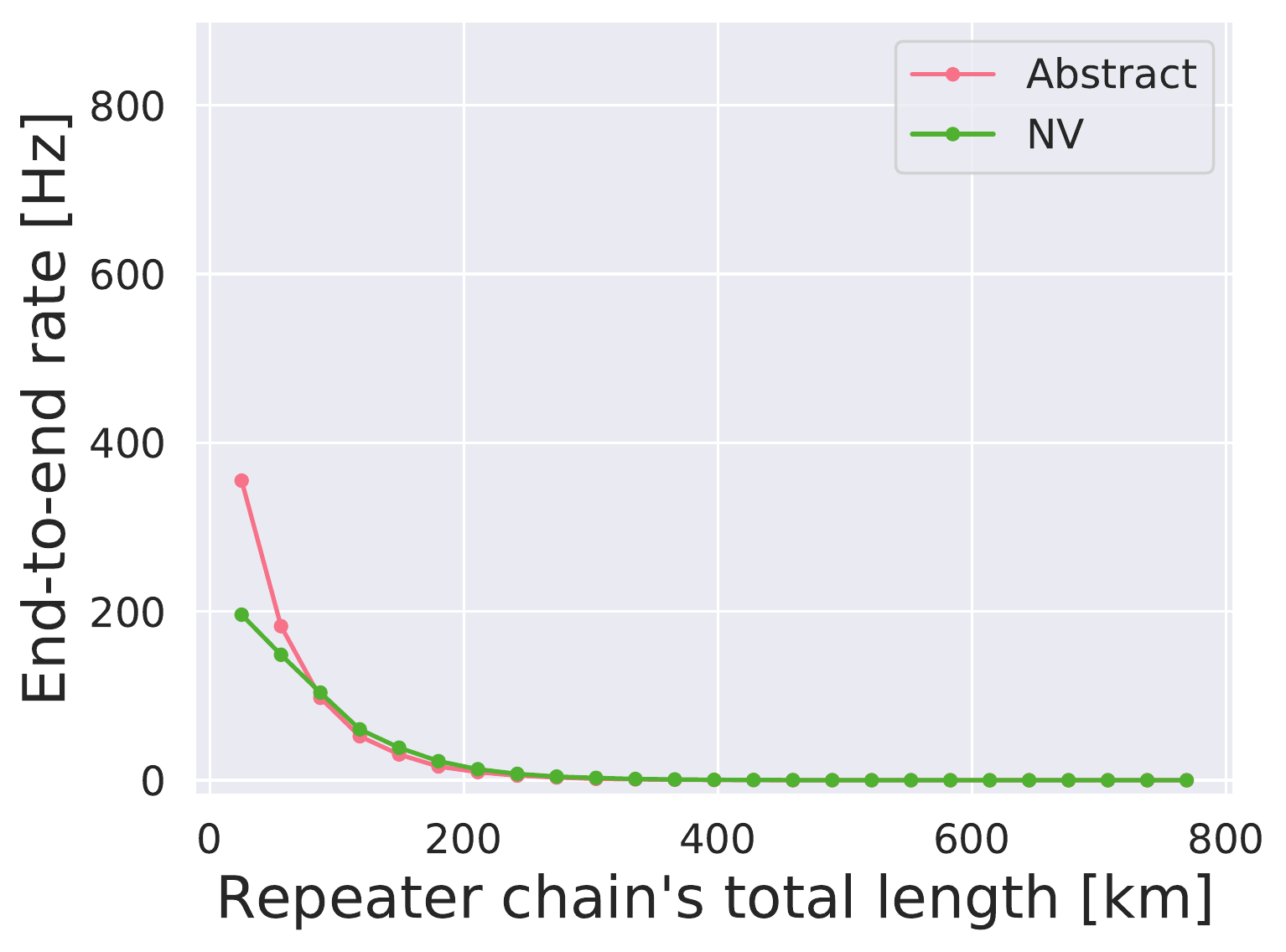}
\caption{Variation of the end-to-end entanglement generation rate in a chain of five equally spaced nodes as the chain's length is varied in the NV model (triangles) and in the abstract model (circles). At short lengths, the rates achieved are higher in the abstract model by a factor of almost $2$. As the distance increases, the two curves overlap.}
\label{fig:validate_abstract_5nodes_rate_distance}
\end{figure}

We turn our attentions now to the other metric of interest, the end-to-end entanglement generation rate. In Figure~\ref{fig:validate_abstract_5nodes_rate_distance}, we plot the end-to-end rate against the total chain length for the same setup in both models.
The behaviour of the two curves is similar, although the rates in the abstract model are significantly higher. We believe that this is due to the fact that, since NV centers have only one communication qubit, they must swap established entanglement from it to a memory qubit as soon as it is generated. This doesn't happen in the abstract model, and thus there is no time spent on swapping the entangled states around qubits, allowing for a higher entanglement generation rate. The difference between the two curves becomes smaller as the distance increases, which could be explained by the fact that at long distances, the majority of the time is spent on generating elementary links, as success probabilities become low. The duration of local node operations become negligible in comparison, and the time taken by internal swaps is not as important in this regime. 

\begin{figure}[!ht]
\centering
\includegraphics[width=\columnwidth]{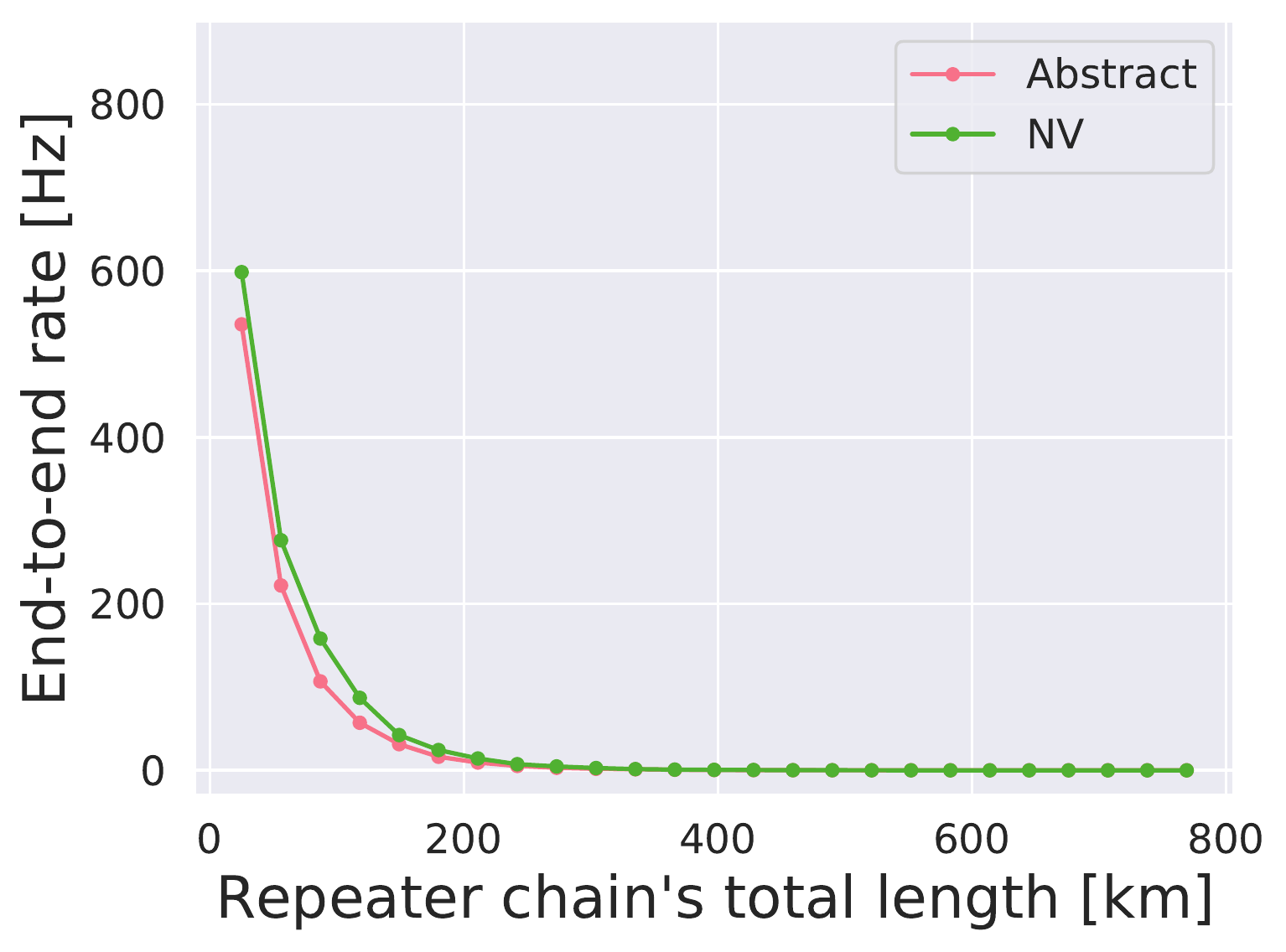}
\caption{Variation of the end-to-end entanglement generation rate in a chain of five equally spaced nodes as the chain's length is varied in the NV model (triangles) and in the abstract model (circles) with instantaneous SWAP gates. The two curves are very close for all simulated chain lengths. The error bars are smaller than the markers.}
\label{fig:Rate_vs_distance_5_nodes_smart_matching_instant_internal_swap}
\end{figure}

In order to verify this, we reran the NV simulation with the internal swap being performed instantly. The results are shown in Figure~\ref{fig:Rate_vs_distance_5_nodes_smart_matching_instant_internal_swap}.
The curves overlap over all distances the simulation covered, corroborating our hypothesis. 

We conclude that the entanglement generation rates attained by the two models are similar across the board, with the the biggest difference, which happens at short internode distances, being a factor of roughly 1.8. At longer distances, the rates are the same up to statistical fluctuations.

\section{Werner Chains}\label{sec:appendixwerner}
In this Appendix we give details about our approach for validating our GA-based optimization approach by applying it to a repeater chain generating Werner states.

The crux of this validation procedure is that we are able to find the optimum value of the cost function by a method other than the GA-based one we proposed. In order to do so, we require closed-form expressions for end-to-end fidelity and entanglement generation rate as functions of the input parameters, elementary link fidelity, success probability and swap quality. 

\begin{figure}[!ht]
\centering
\includegraphics[width=\columnwidth]{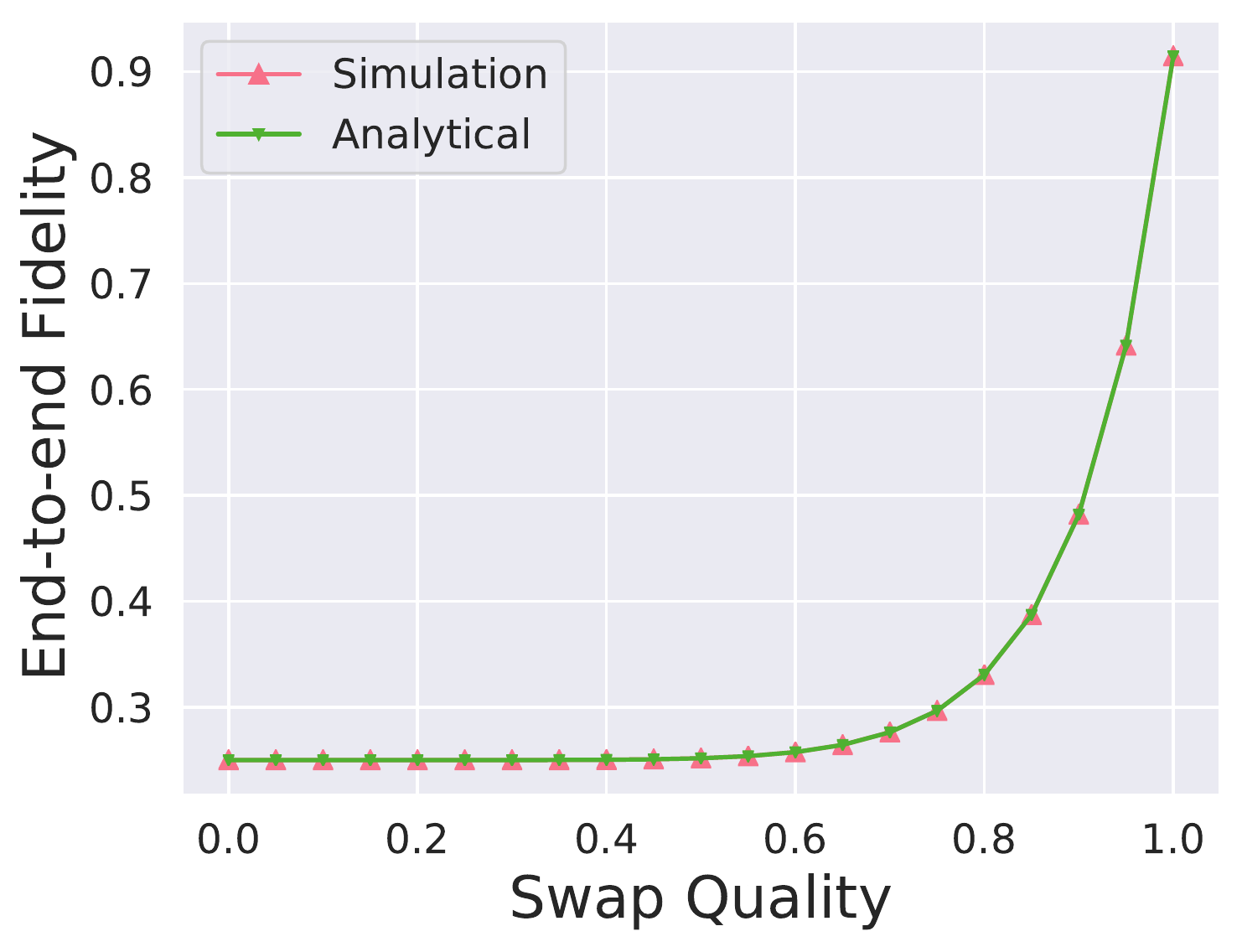}
\caption{End-to-end fidelity of states generated by a chain of ten nodes as the swap quality is varied, for an elementary link fidelity of $0.99$. The analytical and simulation curves perfectly overlap.}
\label{fig:fidelity_comparison_simulation_analytical}
\end{figure}

Consider first, for simplicity, a three-node chain. The nodes establish elementary links whose states are of the form

\begin{equation}
    \rho(x) = x\ket{\psi^+}\bra{\psi^+} + (1-x)\frac{\mathbb{I}}{4},
\end{equation}
where $\ket{\psi^+} = 1/\sqrt{2}(\ket{01} + \ket{10})$ is the ideal Bell state and $\mathbb{I}$ is the identity. $x$ is the Werner parameter and is related to the fidelity $f$ of the Werner state with the ideal Bell state by $f = (1 + 3x)/4$. Performing an ideal BSM on two of these states, both of parameter $x$, results in a Werner state of parameter $x^2$, i.e. the post-BSM state $\rho_{BSM}$ is given by
\begin{equation}
    \rho_{BSM} = x^2\ket{\psi^+}\bra{\psi^+} + (1-x^2)\frac{\mathbb{I}}{4}.
\end{equation}
To simulate a noisy BSM, we then apply noise via two single-qubit depolarizing channels, one on each of the two qubits involved in the BSM. Both of these channels are parametrized by the swap quality $s_q$, as defined in Equation~\eqref{eq:depol_channel}. The resulting Werner state has fidelity $F$ with the ideal Bell state: 
\begin{equation}
    F(f,s_q) = \frac{1}{4} + s_q\left(\frac{1}{2}+\frac{s_q}{4}\right)\left(\frac{4f - 1}{3}\right)^2.
\end{equation}
Iterating this process, one arrives at the following expression for the end-to-end fidelity
\begin{equation}
    F(N,f,s_q) =  \frac{1}{4} + s_q^N\left(\frac{1}{2}+\frac{s_q^N}{4}\right)\left(\frac{4f - 1}{3}\right)^{N+1},
\label{eq:fidelity_analytical}
\end{equation}
where $N$ is the number of repeater nodes in the chain. As a sanity check, we ran simulations of a 10-node chain for a fixed $f$ while varying $s_q$ and compared the obtained end-to-end fidelity with the values obtained with Equation $\eqref{eq:fidelity_analytical}$. These results are shown in Figure~\ref{fig:fidelity_comparison_simulation_analytical}.

An attentive reader might notice that Equation~\eqref{eq:fidelity_analytical} slightly differs from the well-known result first derived in~\cite{dur1999quantum} in how it accounts for the effect of imperfect operations in the end-to-end fidelity. This is due to the fact that we have here parametrized the depolarizing noise in a slightly different manner, through two single-qubit channels.

We shift now our focus to the computation of the end-to-end entanglement generation rate across a 3-node repeater chain. We note that this quantity is simply the inverse of the waiting time, which we denote by $T$. Let us start with the generation of elementary links. Since we model elementary link generation attempts as processes succeeding with a fixed probability $p_{suc}$, $T_0$ is a discrete random variable following a geometric distribution. Its expected value is then given by:

\begin{equation}
    \mathbb{E}(T_0) = \frac{1}{p_{suc}}T_{cycle},
\end{equation}
where $\mathbb{E}$ denotes the expected value and $T_{cycle}$ is the cycle time, i.e. the time a single entanglement generation attempt takes. We consider a sequential repeater chain, i.e. one in which nodes can only attempt entanglement generation with one of their neighbours at a time. Therefore, the end-to-end waiting time is given by:

\begin{equation}
    \mathbb{E}(T) = 2 \mathbb{E}(T_0) +  T_{SWAP},
\end{equation}
where $T_{SWAP}$ is the time an entanglement swap takes. This holds because the repeater node has to generate elementary links with both its neighbours, and it can only start generating the second once it has finished generating the first. Furthermore, after having generated these links, it must swap them. We then define the entanglement rate $R$ as the inverse of the expected waiting time:
\begin{equation}
    R = \frac{1}{\mathbb{E}{\left(T\right)}}.
\label{eq:analyticalrate}
\end{equation}

With Equations~\eqref{eq:fidelity_analytical} and~\eqref{eq:analyticalrate} in hand, we can compute the end-to-end fidelity and entanglement generation rate using only the input parameters $f$, $s_q$ and $p_{suc}$ and the simulation parameters $T_{cycle}$ and $T_{SWAP}$. This implies that we can also directly compute the cost function, as we have analytical expressions for every term appearing in the cost function defined Equation~\eqref{eq:totalcost}. We then used the Basin-Hopping algorithm~\cite{wales1997global} to find the global minimum of this cost function for a target fidelity $f_{min} = 0.6$ and a target entanglement generation rate $r_{min} = 1$ Hz over a chain of equally spaced nodes. We took as baseline values $f_b = s_{q_b} = 0.5$ and $p_{suc_b} = 10^{-10}$. The Basin-Hopping algorithm is available in the SciPy library.

\section{Computing Baseline Values in the Abstract Model}
\subsection{Uniform Spacing}
In order to use a realistic and up to date set of baseline values, we considered the latest results achieved in Ronald Hanson's lab at QuTech, in Delft~\cite{privatecomm}. The values for $T_1$ and $T_2$ can be directly computed from experimental values. The same is true for $s_q$, which can be derived from entanglement swap experiments. This does not hold for the elementary link-related parameters, namely the fidelity $F_{EL}$ and success probability $p_{suc}$. Their values are heavily distance-dependent, and to date entanglement generation experiments using NV centers have only been realized at distances on the single kilometer scale~\cite{hensen2015loophole}. We therefore use instead the model proposed in~\cite{kalb2017entanglement} with the experimental values we obtained from the Hanson group as inputs to compute the baseline values for $F_{EL}$ and $p_{suc}$ for the elementary link lengths we consider. In Table~\ref{tab:nv_values} we list the values used as inputs to the NV model to compute the baseline abstract parameter values. Explaining the physical meaning of each of these parameters would require a detailed exposition of the NV model, which is beyond the scope of this work. This can instead be found in \cite{kalb2017entanglement, dahlberg2019link}. We note that although these parameter values have all been measured in actual laboratory experiments, they are not absolute truths. Different setups might achieve slightly different performances, and even in the same NV center not all nuclear spins are identical nor do they couple in exactly the same way to the electron spin. These nonetheless provide a valuable picture of the current state of the art.

\begin{table}[!ht]
\begin{tabular}{|c|c|}
\hline
Parameter                            & Value              \\ \hline
visibility                           & 0.90               \\ \hline
$\sigma$ phase drift                 & 0.35 rad               \\ \hline
$p_{\text{double excitation}}$       & 0.06              \\ \hline
$p_{\text{electron measure error}}$  & 0.025              \\ \hline
$p_{\text{electron 1 qubit error}}$  & 0.                 \\ \hline
$F_{\text{carbon Z rot}}$      & 0.999            \\ \hline
$F_{\text{EC}}$           & 0.97              \\ \hline
$T_{\text{1 carbon}}$                & 10 h               \\ \hline
$p_{\text{det}}$                     & 0.00013            \\ \hline
$p_{\text{dark count}}$              & $2.5\times10^{-6}$ \\ \hline
N$_{1/e}$ & 1400              \\ \hline
$p_{\text{loss length}}$ & 0.5 dB/km \\ \hline
\end{tabular}
\label{tab:nv_values}
\caption{Values considered for NV model parameters. See e.g. \cite{kalb2017entanglement, dahlberg2019link, rozpkedek2019near} for detailed explanations of parameters.}
\end{table}
The bright state population $\alpha$ is also a required parameter in the model. We chose not to include it in Table~\ref{tab:nv_values} as this parameter is not defined by the quality of the hardware but can instead be chosen. It represents the fraction of the NV electron spin that is in the bright state, i.e. the state that emits photons. It therefore has a direct effect on the success probability of establishing elementary links, as a bigger $\alpha$ results in a higher photon emission probability. On the other hand, increasing $\alpha$ also increases the fraction of terms orthogonal to the Bell basis in the entangled state, decreasing the elementary link fidelity. There is thus a trade-off between elementary link fidelity and success probability when varying an NV center's bright state population~\cite{kalb2017entanglement}. However, in our simplified abstract model we ignore any correlations between parameters, so such a trade-off is not present. We therefore chose to ignore the existence of the trade-off in NV centers when computing the baseline value. Our process for computing these values consisted of performing a parameter scan over $\alpha$ with the NV model and choosing the highest achievable elementary link fidelity and success probability. In practice, this means that the baseline values considered for the elementary link fidelity were obtained with very low values of $\alpha$ and, conversely, the baseline values of the elementary link success probability were obtained with the highest values of $\alpha$. We note that we restricted the parameter scan to the $[0, 0.5]$ interval, because for $\alpha > 0.5$ entanglement is impossible even for perfect parameters. 

Taking all of this into account, we show in Table~\ref{tab:baseline_values} the baseline values we obtained for the abstract model parameters. The distances in the table correspond to the elementary link lengths we considered in the two uniform spacing use cases. 

\begin{table}[!ht]
\begin{tabular}{|c|c|c|c|c|c|}
\hline
 &  $73$ km  & 89 km   & $100$ km   &  $133$km  & $200$ km  \\ \hline
$T_1$ & \multicolumn{5}{c|}{$10$ h} \\ \hline
$T_2$ & \multicolumn{5}{c|}{$4.9$ ms} \\ \hline
$s_q$ & \multicolumn{5}{c|}{$0.8459$} \\ \hline
$F_{EL}$ &  $0.95$  &  $0.94$  &  $0.90$  &  $0.80$  &  $0.52$ \\ \hline
$p_{suc}$ &  $1.3\times10^{-4}$   &  $7.0\times10^{-5}$  & $1.5\times10^{-5}$   &  $2.2\times10^{-6}$  & $9.6\times10^{-8}$  \\ \hline
\end{tabular}
\label{tab:baseline_values}
\caption{Baseline values of the abstract model parameters for the different elementary link lengths considered.}
\end{table}
\label{sec:appendixbaseline}

\subsection{Real Network}
The way we arrive at the baseline values used in this use case is identical to what was described in the previous section, with the exception of $F_{EL}$ and $p_{suc}$. We will now explain why these values must be computed in a different manner, as well as the process we employed to do so.

In order to arrive at realistic baseline values for the network we introduced in Figure~\ref{fig:real_life_network} we used real-life fiber data that was made available to us by SURF. Although we cannot share this data, we used both the physical length of the fibers connecting the locations indicated in the Figure~\ref{fig:real_life_network} and their measured attenuation values. These two quantities then have an impact on the baseline values we consider for $F_{EL}$ and $p_{suc}$, resulting in three different sets of baseline values, one for each of the links in the network. This raises some questions about how the value of the cost function introduced in Equation~\eqref{eq:costfunction} should be computed, as this function takes as input only one set of baseline values and a respective set of improved values. There are multiple ways to address this. We will now explain the approach we took. 

We start by computing four sets of baseline values: one for each of the links in the network plus one at negligible fiber length. By this we mean that the length we use as an input to the model in~\cite{kalb2017entanglement} is such that the impact of losses in the fiber are negligible. The cost associated with a given set of parameters is computed with respect to the set of baseline values at negligible fiber length. One can then think of this set of parameters as the improved parameters at negligible fiber length. In order to obtain the sets of parameters that will be used in our simulation we start by obtaining the improvement factor, defined in Equation~\eqref{eq:improvement_factor}, for each of the parameters. These improvement factors are then applied to the baseline values of each of the links according to Equation~\eqref{eq:progressiveimprovement}. The resulting three sets of values, one for each of the links, are finally the ones fed into our simulation. We reiterate that this process only applies to $F_{EL}$ and $p_{suc}$. The baseline values of the remaining parameters, not being dependent on fiber length, are computed in the same way as described in the previous section. In Table~\ref{tab:baseline_values_real} we present the baseline values we arrived at through the aforementioned process.
\begin{table}[!ht]
\begin{tabular}{|c|c|c|}
\hline
   & $p_{suc}$ & $F_{EL}$ \\ \hline
DH & 0.002588  & 0.9683   \\ \hline
HL & 0.0009187 & 0.9643   \\ \hline
LA & 0.0009082 & 0.9642   \\ \hline
NL & 0.004600  & 0.9698   \\ \hline
\end{tabular}
\label{tab:baseline_values_real}
\caption{Baseline values for the links (DH stands for Delft - The Hague, HL for The Hague - Leiden and LA for Leiden - Amsterdam) and at negligible fiber length (NL).}
\end{table}

\section{Search space reduction using previous runs}
We can use previous optimization runs to limit the search space of new runs and hence increase the probability of a good solution being found. As an example of how this can be done, suppose we have performed an optimization run over a repeater chain of $5$ uniformly spaced nodes spanning some distance $L$. This resulted in a solution that achieves an end-to-end entanglement generation rate of $R = 1$ Hz with an elementary link success probability of $p_{suc_5}$, the subscript being here used to denote the number of nodes in the chain. Say we now want to apply our optimization method to a chain of $7$ uniformly spaced nodes spanning the same distance $L$. As more elementary links need to be established and more entanglement swaps need to be performed, we know with certainty that, in order to achieve the same $R$ a higher elementary link success probability will be needed, i.e. $p_{suc_7} > p_{suc_5}$. We can thus impose a lower bound of $p_{suc_5}$ on the search space, reducing it.

These considerations are easy to make for the case of the elementary link success probability. Since we hold the operation times constant and implement no cut-off, it is the only parameter influencing the end-to-end entanglement generation rate. The same is not true for the other metric of interest, the end-to-end fidelity. As a concrete example, assume that the best solution found for a repeater chain of $5$ uniformly spaced nodes had an elementary link fidelity $F_{EL} = 0.96$ and a swap quality $s_q = 0.98$, resulting in an end-to-end fidelity of $0.75$. One could be inclined to, in a future optimization run, upper bound the search space of $F_{EL}$ by $0.96$ to help lead the algorithm to a solution with an end-to-end fidelity closer to the target value of $0.7$. However, it might be that there is a solution with $F_{EL} > 0.96$ and $s_q < 0.98$ that results in a lower cost function value than any solution with $F_{EL} < 0.96$. Therefore, by imposing this upper bound we could be preventing the algorithm from ever finding the ideal solution.

\end{document}